\documentstyle[preprint,eqsecnum,aps]{revtex}
\tightenlines

\newtheorem{theorem}{Theorem}
\newtheorem{acknowledgement}[theorem]{Acknowledgement}

\begin{document}
\draft
\title{{\Large New solutions of relativistic wave equations in magnetic fields and
longitudinal fields. }}
\author{V.G. Bagrov\thanks{%
On leave from Tomsk State University and Tomsk Institute of High Current
Electronics, Russia}, M.C. Baldiotti\thanks{%
e-mail: baldiott@fma.if.usp.br}, D.M. Gitman\thanks{%
e-mail: gitman@fma.if.usp.br}, and I.V. Shirokov\thanks{%
Omsk State University, Russia; e-mail: shirokov@univer.omsk.su}}
\address{Instituto de F\'{\i}sica, Universidade de S\~ao Paulo,\\
C.P. 66318, 05315-970 S\~ao Paulo, SP, Brasil}
\date{\today}
\maketitle

\begin{abstract}
We demonstrate how one can describe explicitly the present arbitrariness in
solutions of relativistic wave equations in external electromagnetic fields
of special form. This arbitrariness is connected to the existence of a
transformation, which reduces effectively the number of variables in the
initial equations. Then we use the corresponding representations to
construct new sets of exact solutions, which may have a physical interest.
Namely, we present new sets of stationary and nonstationary solutions in
magnetic field and in some superpositions of electric and magnetic fields.
\end{abstract}

\section{Introduction}

Relativistic wave equations (Dirac and Klein-Gordon) provide a basis for
relativistic quantum mechanics and quantum electrodynamics of spinor and
scalar particles \cite{Schwe61}. In relativistic quantum mechanics,
solutions of relativistic wave equations are referred to as one-particle
wave functions of fermions and bosons in external electromagnetic fields. In
quantum electrodynamics, such solutions allow the development of the
perturbation expansion known as the Furry picture, which incorporates the
interaction with the external field exactly, while treating the interaction
with the quantized electromagnetic field perturbatively \cite{FraGiS91}. The
physically most important exact solutions of the Klein-Gordon and the Dirac
equations are: an electron in a Coulomb field, a uniform magnetic field, the
field of a plane wave, the field of a magnetic monopole, the field of a
plane wave combined with a uniform magnetic and electric fields parallel to
the direction of wave propagation, crossed fields, and some simple
one-dimensional electric fields (for a complete review of solutions of
relativistic wave equations see \cite{BagGi90}).

Considering, for example, stationary solutions of relativistic wave
equations, we can see that in the general case, there exist different sets
of stationary solutions for one and the same Hamiltonian. The possibility to
get different sets of stationary states reflects the existence of an
arbitrariness in the solutions of the eigenvalue problem for a Hamiltonian.
Considering nonstationary solutions, we also encounter the possibility of
constructing different complete sets of such solutions. There is no regular
method of describing such an arbitrariness explicitly. Especially in the
presence of an external field the problem appears to be nontrivial.

In the present article we demonstrate how one can describe explicitly the
present arbitrariness in solutions of the relativistic wave equations for
some types of external electromagnetic fields, namely, for uniform magnetic
fields and combination of these fields with some electric fields. This
arbitrariness is connected to the existence of a transformation, which
reduces effectively the number of variables in the initial equations. Then
we use the corresponding representations to construct new sets of exact
solutions, which may have a physical interest. In Sect.II we consider
relativistic wave equations in pure uniform magnetic fields. Here we derive
a representation for the exact solutions, in which the above mentioned
arbitrariness is described explicitly by an arbitrary function. From a
suitable choice of this function, we get both the well-known set of
solutions and new ones. This Section contains the most complete (at the
present) description of the problem of a uniform magnetic field in
relativistic quantum mechanics. Among new sets of solutions there are both
stationary, generalized coherent solutions and nonstationary solutions.
Then, in Sect.III, we consider more complicated configurations of external
electromagnetic fields, namely, longitudinal electromagnetic fields. Here we
describe all the arbitrariness in the solutions, and on this base present
various sets of new exact solutions. In Sect.IV we interpret the above
results from the point of view of the general theory of differential
equations.

\section{Uniform magnetic field}

\subsection{Arbitrariness in solutions of relativistic wave equations.}

Consider a uniform magnetic field ${\bf H=}\left( 0,0,H\right) $\ directed
along the $x^{3}$ axis ($H>0)$. The electromagnetic potentials are chosen in
the symmetric gauge

\begin{equation}
A_{0}=A_{3}=0,\,\;A_{1}=\frac{1}{2}Hx^{2},\;\,A_{2}=-\frac{1}{2}Hx^{1}\;.
\label{2.1}
\end{equation}
We write the Klein-Gordon and the Dirac equations in the form 
\begin{eqnarray}
&&{\cal K}\Psi =0,\quad \hbar ^{2}{\cal K}={\cal P}^{2}-m_{0}^{2}c^{2},\;%
{\cal P}_{\mu }=i\hbar \partial _{\mu }-\frac{e}{c}A_{\mu }\,,  \nonumber \\
&&{\cal D}{\bf \Psi }=0\;,\quad \hbar {\cal D}=\gamma ^{\mu }{\cal P}_{\mu
}-m_{0}c\;.  \label{2.2}
\end{eqnarray}
Here $e=-|e|$ and $\gamma $-matrices are chosen in the standard
representation \cite{BagGi90}.

In the field under consideration, the operators ${\cal P}_{0}$ and ${\cal P}%
_{3}$ are mutually commuting integrals of motion, $\left[ {\cal K},{\cal P}%
_{0}\right] =\left[ {\cal K},{\cal P}_{3}\right] =\left[ {\cal D},{\cal P}%
_{0}\right] =\left[ {\cal D},{\cal P}_{3}\right] =\left[ {\cal P}_{0},{\cal P%
}_{3}\right] =0$.

In the case of the Klein-Gordon equation, the operator $L_{z},$ 
\begin{equation}
L_{z}=i\hbar \left( x^{2}\partial _{1}-x^{1}\partial _{2}\right) ,\;\left[
L_{z},{\cal P}_{0}\right] =\left[ L_{z},{\cal P}_{3}\right] =\left[ {\cal K}%
,L_{z}\right] =0\;,  \label{2.3}
\end{equation}
can be included (together with ${\cal P}_{0}$ and ${\cal P}_{3}$) in the
complete set of integrals of motion, whereas for the Dirac equation case,
the operator $J_{z}$, 
\begin{equation}
J_{z}=L_{z}+\frac{\hbar }{2}\Sigma _{3},\;\left[ J_{z},{\cal P}_{0}\right] =%
\left[ J_{z},{\cal P}_{3}\right] =\left[ {\cal D},J_{z}\right] =0\,,\;
\label{2.4}
\end{equation}
can be included (together with ${\cal P}_{0}$ and ${\cal P}_{3}$) in the
complete set of integrals of motion. Here $\Sigma _{3}={\rm diag}\left(
\sigma _{3},\sigma _{3}\right) $ .

We are going to use dimensionless coordinates $-\infty <x<\infty ,\;-\infty
<y<\infty $ or $0\leq \rho <\infty ,\;0\leq \varphi <2\pi $ defined by the
relations 
\begin{eqnarray}
&&\sqrt{\frac{\gamma }{2}}x^{1}=x=\sqrt{\rho }\cos \varphi ,\quad \sqrt{%
\frac{\gamma }{2}}x^{2}=y=\sqrt{\rho }\sin \varphi \;,\quad \gamma =\frac{%
\left| e\right| H}{c\hbar }>0\;,  \nonumber \\
&&dx^{1}dx^{2}=\frac{2}{\gamma }dx\,dy=\frac{1}{\gamma }d\rho \,d\varphi
,\quad x+iy=\sqrt{\rho }\exp i\varphi \;.  \label{2.5}
\end{eqnarray}
It is useful to introduce the operators $a_{1},a_{2},a_{1}^{+},a_{2}^{+}$ , 
\begin{eqnarray}
a_{2} &=&\frac{1}{\sqrt{2\gamma }\hbar }\left[ {\cal P}_{2}-i{\cal P}%
_{1}+\hbar \gamma \left( x^{1}+ix^{2}\right) \right] =\frac{1}{2}\left(
x+iy+\partial _{x}+i\partial _{y}\right) =\frac{e^{i\varphi }}{2\sqrt{\rho }}%
\left( \rho +i\partial _{\varphi }+2\rho \partial _{\rho }\right) , 
\nonumber \\
a_{2}^{+} &=&\frac{1}{\sqrt{2\gamma }\hbar }\left[ {\cal P}_{2}+i{\cal P}%
_{1}+\hbar \gamma \left( x^{1}-ix^{2}\right) \right] =\frac{1}{2}\left(
x-iy-\partial _{x}+i\partial _{y}\right) \,=\frac{e^{-i\varphi }}{2\sqrt{%
\rho }}\left( \rho +i\partial _{\varphi }-2\rho \partial _{\rho }\right) . 
\nonumber \\
a_{1} &=&-\frac{1}{\sqrt{2\gamma }\hbar }\left( i{\cal P}_{1}+{\cal P}%
_{2}\right) =\frac{1}{2}\left( x-iy+\partial _{x}-i\partial _{y}\right) =%
\frac{e^{-i\varphi }}{2\sqrt{\rho }}\left( \rho -i\partial _{\varphi }+2\rho
\partial _{\rho }\right) ,  \nonumber \\
a_{1}^{+} &=&\frac{1}{\sqrt{2\gamma }\hbar }\left( i{\cal P}_{1}-{\cal P}%
_{2}\right) =\frac{1}{2}\left( x+iy-\partial _{x}-i\partial _{y}\right) =%
\frac{e^{i\varphi }}{2\sqrt{\rho }}\left( \rho -i\partial _{\varphi }-2\rho
\partial _{\rho }\right) .  \label{2.7}
\end{eqnarray}
They obey the commutation relations 
\begin{equation}
\left[ a_{k},a_{s}^{+}\right] =\delta _{k,s\,},\quad \left[ a_{k},a_{s}%
\right] =\left[ a_{k}^{+},a_{s}^{+}\right] =0\,,\quad k,s=1,2\;.
\label{2.10}
\end{equation}
Thus, we can interpret these operators as creation and annihilation ones.
One can also find the following relations 
\begin{eqnarray}
&&{\cal P}_{1}^{2}+{\cal P}_{2}^{2}=\hbar ^{2}\gamma \left(
a_{1}a_{1}^{+}+a_{1}^{+}a_{1}\right) =2\hbar ^{2}\gamma {\cal N}+\hbar
^{2}\gamma \;,  \nonumber \\
&&L_{z}=\hbar \left( {\cal N}-a_{2}^{+}a_{2}\right) \,,\quad {\cal N}%
=a_{1}^{+}a_{1}\;.  \label{2.11}
\end{eqnarray}
Then the Klein-Gordon and the Dirac operators can be written as 
\begin{eqnarray}
{\cal K} &=&\hbar ^{-2}\left( {\cal P}_{0}^{2}-{\cal P}_{3}^{2}\right)
-2\gamma {\cal N}-\gamma -m^{2},\quad m=\frac{m_{0}c}{\hbar }\,,  \nonumber
\\
{\cal D} &=&\hbar ^{-1}\left( \gamma ^{0}{\cal P}_{0}+\gamma ^{3}{\cal P}%
_{3}\right) -\sqrt{\frac{\gamma }{2}}\left[ \left( \gamma ^{2}-i\gamma
^{1}\right) a_{1}+\left( \gamma ^{2}+i\gamma ^{1}\right) a_{1}^{+}\right]
-m\,.  \label{2.12}
\end{eqnarray}

The operator ${\cal N}$ commutes with ${\cal P}_{0},$\ ${\cal P}%
_{3},\;L_{z}, $ plus it is an integral of motion in the case of the
Klein-Gordon equation. Its generalization for the Dirac equation has the
form ${\cal N}_{D}={\cal N}+\frac{1}{2}\Sigma _{3}$.

One ought to remark that the operators ${\cal K}$ and ${\cal D}$ do not
contain the operators $a_{2}^{+},$ $a_{2}.$ Thus, the latter operators are
integrals of motion, which commute with ${\cal N}$, ${\cal N}_{D},{\cal P}%
_{0}$, ${\cal P}_{3},\;$but do not commute with $L_{z}$ and $J_{z}$ .

The operators of creation and annihilation with different numbers commute.
One can find a representation in which these operators are acting on
different variables. To this end, we present the wave functions from (\ref
{2.2}) in the following form (we make a Fourier transform in the variable $y$
only, and call such a representation the semi-momentum representation) 
\begin{equation}
\Psi (x,y)=\frac{1}{\sqrt{2\pi }}\int_{-\infty }^{\infty }e^{iky}\widetilde{%
\Psi }(x,k)\;dk\;.  \label{2.13}
\end{equation}
Of course the functions $\Psi $ and $\tilde{\Psi}$ depend on the variables $%
x^{2}$ and $x^{3}$ as well, but we do not indicate this dependence
explicitly. In terms of $\tilde{\Psi}$ the multiplication and
differentiation have the form $y\rightarrow i\partial _{k},\quad i\partial
_{y}\rightarrow -k$. Then, the expressions for the creation and annihilation
operators in the semi-momentum representation take the form 
\begin{eqnarray}
&&2a_{1}=x+k+\partial _{x}+\partial _{k},\quad 2a_{1}^{+}=x+k-\partial
_{x}-\partial _{k}\;,  \nonumber \\
&&2a_{2}=x-k+\partial _{x}-\partial _{k},\quad 2a_{2}^{+}=x-k-\partial
_{x}+\partial _{k}\;.  \label{2.14}
\end{eqnarray}
Now we pass from $x,\,k$ to new variables $\xi ,\,\eta $, 
\begin{equation}
\sqrt{2}\xi =x+k,\quad \sqrt{2}\eta =x-k,\quad \sqrt{2}x=\xi +\eta ,\quad 
\sqrt{2}k=\xi -\eta \;.  \label{2.15}
\end{equation}
Then the creation and annihilation operators can be written as

\begin{equation}
\sqrt{2}a_{1}=\xi +\partial _{\xi }\,,\quad \sqrt{2}a_{1}^{+}=\xi -\partial
_{\xi }\,,\;\;\sqrt{2}a_{2}=\eta +\partial _{\eta }\,,\quad \sqrt{2}%
a_{2}^{+}=\eta -\partial _{\eta }\;.  \label{2.16}
\end{equation}
In the new variables, 
\begin{equation}
2{\cal N}=\xi ^{2}-\partial _{\xi }^{2}-1\;,  \label{2.17}
\end{equation}
and the Klein-Gordon and the Dirac operators read 
\begin{eqnarray}
{\cal K} &=&\hbar ^{-2}\left( {\cal P}_{0}^{2}-{\cal P}_{3}^{2}\right)
+\gamma \left( \partial _{\xi }^{2}-\xi ^{2}\right) -m^{2}\;,  \nonumber \\
{\cal D} &=&\hbar ^{-1}\left( \gamma ^{0}{\cal P}_{0}+\gamma ^{3}{\cal P}%
_{3}\right) -\sqrt{\gamma }\left( \gamma ^{2}\xi -i\gamma ^{1}\partial _{\xi
}\right) -m\;.  \label{2.18}
\end{eqnarray}
One can see that the latter operators do not contain the variable $\eta $.
Notice that both operators $L_{z}$ and $J_{z}$ contain variables $\xi ,\eta
. $ For example, 
\begin{equation}
2L_{z}=\xi ^{2}-\partial _{\xi }^{2}-\eta ^{2}+\partial _{\eta }^{2}\;.
\label{2.19}
\end{equation}

The integration over $k$ in (\ref{2.13}) can be replaced by an integration
over $\eta ,$ 
\begin{equation}
\Psi \left( x,y\right) =\frac{e^{ixy}}{\sqrt{\pi }}\int\limits_{-\infty
}^{\infty }e^{-i\sqrt{2}y\eta }\widetilde{\Psi }\left( \xi ,\eta \right)
\;d\eta ,\quad \xi =\sqrt{2}x-\eta \;.  \label{2.20}
\end{equation}
Besides, one can write 
\begin{equation}
\left( \Psi ,\Phi \right) =\int\limits_{-\infty }^{\infty
}dx\;\int\limits_{-\infty }^{\infty }dy\;\Psi ^{\ast }\left( x,y\right) \Phi
\left( x,y\right) =\left( \widetilde{\Psi },\widetilde{\Phi }\right)
=\int\limits_{-\infty }^{\infty }d\xi \int\limits_{-\infty }^{\infty }d\eta
\;\widetilde{\Psi }^{\ast }\left( \xi ,\eta \right) \widetilde{\Phi }\left(
\xi ,\eta \right) \;.  \label{2.21}
\end{equation}

The independence of the operators (\ref{2.18}) on the variable $\eta $ will
allow us to separate explicitly the functional arbitrariness in the
solutions (\ref{2.20}), as will be seen below.

\subsection{Stationary states}

Known sets of stationary solutions in a uniform magnetic field (that were
found in the first works \cite{Rabi28,Page30,Pless30,Huff31,Saute31}) are
eigenfunctions of the operators ${\cal P}_{0},{\cal P}_{3},{\cal N}$ \ in
the scalar case and of the operators ${\cal P}_{0},\,{\cal P}_{3},\,{\cal N}%
_{D}$ in the spinor case. Thus for scalar wave functions $\Psi $ we have the
conditions 
\begin{equation}
{\cal P}_{0}\Psi =\hbar k_{0}\Psi ,\quad {\cal P}_{3}\Psi =\hbar k_{3}\Psi
\;,\;{\cal N}\Psi =n\Psi ,\;n=0,1,2,\ldots ,  \label{3.1}
\end{equation}
and for Dirac wave functions ${\bf \Psi }$ the conditions 
\begin{equation}
{\cal P}_{0}{\bf \Psi }=\hbar k_{0}{\bf \Psi },\quad {\cal P}_{3}{\bf \Psi }%
=\hbar k_{3}{\bf \Psi },\quad {\cal N}_{D}{\bf \Psi }=\left( n-\frac{1}{2}%
\right) {\bf \Psi },\quad n=0,1,2,\ldots \,.  \label{3.2}
\end{equation}

Consider first the scalar case. It follows from (\ref{2.18}) that 
\begin{equation}
k_{0}^{2}=m^{2}+\gamma +k_{3}^{2}+2\gamma n=m^{\ast \;2}+k_{3}^{2}+2\gamma
n,\quad m^{\ast \;2}=m^{2}+\gamma \;,  \label{3.3}
\end{equation}
and 
\begin{equation}
\Psi _{n,k_{3}}\left( x^{\mu }\right) =N\exp \left(
-ik_{0}x^{0}-ik_{3}x^{3}\right) \Psi _{n}\left( x,y\right) \,.  \label{3.4}
\end{equation}
Here $N$ is a normalization factor. In the semi-momentum representation (\ref
{2.13}) the function $\Psi _{n}\left( x,y\right) $ has the following image 
\begin{equation}
\widetilde{\Psi }_{n}\left( \xi ,\eta \right) =U_{n}\left( \xi \right) \Phi
\left( \eta \right) ,\quad \xi =\sqrt{2}x-\eta \;.  \label{3.5}
\end{equation}
Here Eqs. (\ref{3.1}), (\ref{2.17}) were used. $U_{n}\left( \xi \right) $
are Hermit functions; they are related to the corresponding polynomials $%
H_{n}\left( \xi \right) $ as $U_{n}\left( x\right) =\left( 2^{n}n!\sqrt{\pi }%
\right) ^{-\frac{1}{2}}\exp \left( -x^{2}/2\right) H_{n}\left( x\right) $ 
\cite{GraRy94}. The function $\Phi \left( \eta \right) $ is arbitrary. The
functions $\Psi _{n}\left( x,y\right) $ from (\ref{3.4}) obey the relations 
\begin{eqnarray}
&&a_{1}\Psi _{n}=\sqrt{n}\Psi _{n-1},\quad a_{1}^{+}\Psi _{n}=\sqrt{n+1}\Psi
_{n+1}\,,\;\;\Psi _{n}\left( x,y\right) =\frac{\left( a_{1}^{+}\right) ^{n}}{%
\sqrt{\Gamma \left( n+1\right) }}\Psi _{0}\left( x,y\right) \;,  \label{3.7}
\\
&&\Psi _{0}\left( x,y\right) =\pi ^{-\frac{3}{4}}\exp \left(
-x^{2}+ixy\right) \int\limits_{-\infty }^{\infty }d\eta \;\exp \left[ -\frac{%
\eta ^{2}}{2}+\sqrt{2}\eta \left( x-iy\right) \right] \Phi \left( \eta
\right) \;.  \label{3.8}
\end{eqnarray}

Dirac wave functions are of the form ${\bf \Psi }_{n,k_{3}}(x^{\mu })=N\exp
\left( -ik_{0}x^{0}-ik_{3}x^{3}\right) {\bf \Psi }_{k_{3},n}\left(
x,y\right) $ with bispinors ${\bf \Psi }_{k_{3},n}\left( x,y\right) $ having
the structure

\begin{equation}
{\bf \Psi }_{n,k_{3}}^{T}(x,y)=\left( c_{1}\Psi _{n-1}(x,y),\;ic_{2}\Psi
_{n}(x,y),\;c_{3}\Psi _{n-1}(x,y),\;ic_{4}\Psi _{n}(x,y)\right) \;.
\label{3.10}
\end{equation}
The functions $\Psi _{n}(x,y)$ are defined by the relations (\ref{2.20}), (%
\ref{3.5}), whereas the constant bispinor $C$ (with the elements $c_{k})$
obeys an algebraic system of equations 
\begin{equation}
AC=0,\quad A=\gamma ^{0}k_{0}+\gamma ^{3}k_{3}-\sqrt{2\gamma n}\gamma
^{1}-m\;.  \label{3.11}
\end{equation}
The condition $\det A=(k_{0}^{2}-k_{3}^{2}-2\gamma n-m^{2})^{2}=0$ results
in an equation which is an analog of (\ref{3.3}), 
\begin{equation}
k_{0}^{2}=k_{3}^{2}+2\gamma n+m^{2}\;.  \label{3.13}
\end{equation}
Since the rank of the matrix $A$ is equal to $2,$ a general solution of (\ref
{3.11}) has the form 
\begin{equation}
C=\left( 
\begin{array}{c}
(k_{0}+m)v \\ 
(\sqrt{2\gamma n}\sigma _{1}-k_{3}\sigma _{3})v
\end{array}
\right) ,\quad C^{+}C=2k_{0}(k_{0}+m)v^{+}v\;,  \label{3.14}
\end{equation}
where $v$ is an arbitrary constant bispinor and $\sigma $ are Pauli
matrices. We can specify $v$ selecting a spin integral of motion (see \cite
{BagGi90}). The state $n=0$ is a special case. Here we must set $%
c_{1}=c_{3}=0,$ that corresponds to the choice $v^{T}=\left(
0,\;c_{2}\right) ,\quad c_{2}\neq 0$. The latter means that $\Sigma _{3}\Psi
_{D}=-\Psi _{D}$. Thus, for $n=0$, the electron spin can only point to the
direction opposite to the magnetic field.

Expressions for $\Psi _{n}(x,y)$ in the semi-momentum representation contain
explicitly a functional arbitrariness, which means that every energy level
is infinitely degenerated.\ Let us demand that the scalar and spinor wave
functions be eigenvectors of the operators $L_{z}$ and $J_{z}$ respectively.
According to (\ref{2.4}) and (\ref{2.11}) that means that the functions $%
\Psi _{n}(x,y)$ have to obey an additional condition 
\begin{eqnarray}
&&a_{2}^{+}a_{2}\Psi _{n}(x,y)=s\Psi _{n}(x,y),\quad s=0,1,2,\ldots \;\;, 
\nonumber \\
&&L_{z}=\hbar \left( n-s\right) =\hbar l,\quad l=n-s,\quad n\geq l>-\infty
,\quad J_{z}=\hbar \left( l-\frac{1}{2}\right) \;.  \label{3.17}
\end{eqnarray}
This condition defines the function $\Phi (\eta )$ according to (\ref{2.16}%
), $a_{2}^{+}a_{2}\Phi _{s}(\eta )=s\Phi _{s}(\eta )$, therefore $\Phi
_{s}(\eta )=U_{s}(\eta )$. Substituting this result into (\ref{3.5}) and
into (\ref{2.20}), and doing the integral over $\eta ,$ we find in the
coordinate representation, 
\begin{equation}
\Psi _{n,s}(x,y)=\frac{(-1)^{n}}{\sqrt{2\pi }}e^{il\varphi }I_{s,n}(\rho )=%
\frac{(-1)^{n}}{\sqrt{2\pi }}\left( \frac{x+iy}{x-iy}\right) ^{\frac{n-s}{2}%
}I_{s,n}(x^{2}+y^{2})\;.  \label{3.19}
\end{equation}
Here $I_{m,n}\left( x\right) $ are Laguerre functions, which are connected
to the corresponding polynomials $L_{n}^{\alpha }\left( x\right) $ by the
relations (see \cite{GraRy94}) $I_{m,n}\left( x\right) =\left( \Gamma \left(
n+1\right) /\Gamma \left( m+1\right) \right) ^{1/2}e^{-\frac{x}{2}}x^{\frac{%
\alpha }{2}}L_{n}^{\alpha }\left( x\right) ,\quad \alpha =m-n$. The states (%
\ref{3.19}) were first obtained in the works \cite
{Rabi28,Page30,Pless30,Huff31,Saute31}. Besides (\ref{3.7}) and (\ref{3.8}),
the functions (\ref{3.19}) obey the following relations as well 
\begin{eqnarray}
&&a_{2}\Psi _{n,s}=\sqrt{s}\Psi _{n,s-1},\quad a_{2}^{+}\Psi _{n,s}=\sqrt{s+1%
}\Psi _{n,s+1}\;,  \nonumber \\
&&\Psi _{n,s}=\frac{(a_{1}^{+})^{n}(a_{2}^{+})^{s}}{\sqrt{\Gamma (n+1)\Gamma
(s+1)}}\Psi _{0,0}\;,\;\;\Psi _{0,0}(x,y)=\frac{1}{\sqrt{\pi }}\exp \left[ -%
\frac{1}{2}(x^{2}+y^{2})\right] =\frac{e^{-\frac{\rho }{2}}}{\sqrt{\pi }}\;.
\label{3.21}
\end{eqnarray}

Below we are going to find new sets of solutions imposing complementary
conditions different from (\ref{3.17}). This results in a different form for
the function $\Phi (\eta ).$

Taking into account that the operators $a_{2}^{+},\,a_{2}$ are integrals of
motion, we may construct stationary states, which are eigenvectors of a
linear combination $A_{2}^{\alpha ,\beta }$ of these operators, 
\begin{equation}
A_{2}^{\alpha ,\beta }=\alpha a_{2}+\beta a_{2}^{+}\;.  \label{3.22}
\end{equation}
Here $\alpha $,$\,\beta $ are arbitrary complex numbers. One has to
distinguish here three nonequivalent cases:

If $|\alpha |^{2}<|\beta |^{2}$, then do not exist any normalizable
eigenvectors of the operator (\ref{3.22}) . We are not going to consider
such case.

If $|\alpha |^{2}=|\beta |^{2},$ then $A_{2}^{\alpha ,\beta }$ is, in fact,
reduced to a Hermitian operator 
\begin{equation}
A_{2}^{\mu }=\mu a_{2}+\mu ^{\ast }a_{2}^{+},\quad A_{2}^{+\mu }=A_{2}^{\mu
},\quad \mu \neq 0\;,  \label{3.23}
\end{equation}
where $\mu $ is an arbitrary complex number.

If $|\alpha |^{2}>|\beta |^{2},$ then without loss of generality we can
assume that operators $\ A_{2}^{\alpha ,\beta }$ have the form 
\begin{equation}
A_{2}^{\alpha ,\beta }=\alpha a_{2}+\beta a_{2}^{+},\quad \left| \alpha
\right| ^{2}-\left| \beta \right| ^{2}=1,\quad \left[ A_{2}^{\alpha ,\beta
},\,A_{2}^{+\alpha ,\beta }\right] =1.  \label{3.24}
\end{equation}
Then $\,A_{2}^{+\alpha ,\beta }$, $\,A_{2}^{\alpha ,\beta }$ are creation
and annihilation operators, which are related to $a_{2}^{+},\,a_{2}$ by a
canonical transformation 
\begin{equation}
a_{2}=\alpha ^{\ast }A_{2}^{\alpha ,\beta }-\beta A_{2}^{+\;\alpha ,\beta
},\quad a_{2}^{+}=\alpha A_{2}^{+\;\alpha ,\beta }-\beta ^{\ast
}A_{2}^{\alpha ,\beta }\,.  \label{3.25}
\end{equation}
Consider eigenvectors of the operator (\ref{3.23}), i.e., $A_{2}^{\mu }\Psi
_{n,z}^{\mu }(x,y)=z\Psi _{n,z}^{\mu }(x,y),\quad z=z^{\ast }$. This
equation results in the equation $A_{2}^{\mu }\Phi _{z}^{\mu }\left( \eta
\right) =z\Phi _{z}^{\mu }\left( \eta \right) $ for the function $\Phi (\eta
)$. Taking into account (\ref{2.16}), one can find that solutions of the
latter equation are 
\begin{eqnarray}
&&\Phi _{z}^{\mu }\left( \eta \right) =\left[ \frac{\mu }{\sqrt{2}\pi \left|
\mu \right| \left( \mu -\mu ^{\ast }\right) }\right] ^{\frac{1}{2}}\,\exp
Q_{1},  \nonumber \\
&&4\left( \mu -\mu ^{\ast }\right) Q_{1}=-2\left( \mu +\mu ^{\ast }\right)
\eta ^{2}+4\sqrt{2}z\eta -z^{2}\left( \mu +\mu ^{\ast }\right) \left| \mu
\right| ^{-2}\;.  \label{3.28}
\end{eqnarray}
These solutions obey the orthonormality and completeness relations 
\begin{equation}
\int_{-\infty }^{\infty }\Phi _{z^{\prime }}^{\ast \;\mu }(\eta )\Phi
_{z}^{\mu }(\eta )\;d\eta =\delta (z-z^{\prime })\,,\;\int_{-\infty
}^{\infty }\Phi _{z}^{\ast \;\mu }(\eta ^{\prime })\Phi _{z}^{\mu }(\eta
)\;dz=\delta (\eta -\eta ^{\prime })\,.  \label{3.30}
\end{equation}
Their overlapping has the form 
\begin{eqnarray}
&&R^{\mu \prime ,\mu }\left( z^{\prime },z\right) =\int\limits_{-\infty
}^{\infty }\Phi _{z^{\prime }}^{\ast \;\mu \prime }(\eta )\Phi _{z}^{\ast
\;\mu }(\eta )\;d\eta =N_{1}\exp \left[ \frac{Q_{2}}{4\left( \mu ^{\prime
}\mu ^{\ast }-\mu \mu ^{^{\prime }\ast }\right) }\right] \;,  \nonumber \\
&&N_{1}^{2}=\frac{\mu ^{^{\prime }\ast }\mu }{2\pi ^{2}\left| \mu ^{\prime
}\right| \left| \mu \right| \left( \mu \mu ^{^{\prime }\ast }-\mu ^{\prime
}\mu ^{\ast }\right) }\;,  \nonumber \\
&&Q_{2}=\left( z\sqrt{\frac{\mu ^{\prime }}{\mu }}-z^{\prime }\sqrt{\frac{%
\mu }{\mu ^{\prime }}}\right) ^{2}+\left( z\sqrt{\frac{\mu ^{^{\prime }\ast }%
}{\mu ^{\ast }}}-z^{\prime }\sqrt{\frac{\mu ^{\ast }}{\mu ^{^{\prime }\ast }}%
}\right) ^{2}\;.  \label{3.31}
\end{eqnarray}
It defines the mutual decomposition 
\begin{equation}
\Phi _{z}^{\mu }\left( \eta \right) =\int\limits_{-\infty }^{\infty }\Phi
_{z^{\prime }}^{\mu \prime }\left( \eta \right) R^{\mu ^{\prime },\mu
}\left( z^{\prime },z\right) \;dz^{\prime }\;.  \label{3.32}
\end{equation}
The coordinate representation (\ref{2.20}) for the solutions under
consideration has the form 
\begin{eqnarray}
&&\Psi _{n,z}^{\mu }\left( x,y\right) =\left( \sqrt{2}\pi \left| \mu \right|
\right) ^{-\frac{1}{2}}\left( \frac{\mu ^{\ast }}{\mu }\right) ^{\frac{n}{2}%
}U_{n}\left( p_{1}\right) \exp iQ_{3}\;,  \nonumber \\
&&4\left| \mu \right| ^{2}Q_{3}=\left[ i\left( \mu ^{\ast }-\mu \right)
x+\left( \mu +\mu ^{\ast }\right) y\right] \left[ \left( \mu +\mu ^{\ast
}\right) x+i\left( \mu -\mu ^{\ast }\right) y-2z\right] \;,  \nonumber \\
&&\sqrt{2}\left| \mu \right| p_{1}=\left( \mu +\mu ^{\ast }\right) x+i\left(
\mu -\mu ^{\ast }\right) y-z\;.  \label{3.33}
\end{eqnarray}
Their scalar product (\ref{2.21}) reads $\left( \Psi _{n^{\prime },z^{\prime
}}^{\mu },\Psi _{n,z}^{\mu }\right) =\delta _{n,n^{\prime }}\delta
(z-z^{\prime })$. The relation (\ref{3.32}) results into the following
decomposition in the coordinate representation 
\begin{equation}
\Psi _{n,z}^{\mu }\left( x,y\right) =\int\limits_{-\infty }^{\infty }\Psi
_{n,z^{\prime }}^{\mu ^{\prime }}\left( x,y\right) R^{\mu ^{\prime },\mu
}\left( z^{\prime },z\right) \;dz^{\prime }\;.  \label{3.35}
\end{equation}
In particular, in the cases of real or pure imaginary $\mu $, such wave
functions were known before \cite{BagGi90}.

Consider eigenvectors of the operator (\ref{3.24}), i.e., $A_{2}^{\alpha
,\beta }\Psi _{n,z}^{\alpha ,\beta }\left( x,y\right) =z\Psi _{n,z}^{\alpha
,\beta }\left( x,y\right) $ where $z$ is a complex number. In fact, we get
coherent (squeezed) stationary states. They are labeled by $z$ and by two
complex parameters $\alpha ,\,\beta ,$ which are related by the condition (%
\ref{3.24}). In the semi-momentum representation the above equation is
reduced to the one 
\begin{equation}
A_{2}^{\alpha ,\beta }\Phi _{z}^{\alpha ,\beta }\left( \eta \right) =z\Phi
_{z}^{\alpha ,\beta }\left( \eta \right) \;.  \label{3.37}
\end{equation}

It is well known that such solutions form a complete (overcomplete) set at
any fixed $\alpha ,\,\beta .$ Solutions within each set\ are not orthogonal.
One can use these functions to construct a orthogonal set of solutions.

Since the operators $A_{2}^{+\;\alpha ,\beta }\,,\,A_{2}^{\alpha ,\beta }$
are integrals of motion (both for the Klein-Gordon equation and for the
Dirac equation), they are symmetry operators for the equations. The action
of these operators on a solution provides again a solution. For example,
applying the operators $\left( \Gamma \left( 1+s\right) \right)
^{-1/2}\left( A_{2}^{+\;\alpha ,\beta }-z^{\ast }\right) ^{s},\quad
s=0,1,2,\ldots $ to normalized solutions of the equation (\ref{3.37}), we
get normalized solutions labeled by the index $s.$ These new solutions are
orthogonal with respect to $s$, 
\begin{eqnarray}
&&\Phi _{s,z}^{\alpha ,\beta }\left( \eta \right) =\frac{\left(
A_{2}^{+\;\alpha ,\beta }-z^{\ast }\right) ^{s}}{\sqrt{\Gamma \left(
1+s\right) }}\Phi _{z}^{\alpha ,\beta }\left( \eta \right) ,\quad \Phi
_{0,z}^{\alpha ,\beta }\left( \eta \right) =\Phi _{z}^{\alpha ,\beta }\left(
\eta \right) \;,  \nonumber \\
&&\int\limits_{-\infty }^{\infty }\Phi _{s^{\prime },z}^{\ast \;\alpha
,\beta }\left( \eta \right) \Phi _{s,z}^{\alpha ,\beta }\left( \eta \right)
\;d\eta =\delta _{s,s^{\prime }}\int\limits_{-\infty }^{\infty }\left| \Phi
_{z}^{\alpha ,\beta }\left( \eta \right) \right| ^{2}\;d\eta \;.
\label{3.38}
\end{eqnarray}
We call such states generalized squeezed coherent states. It is possible to
get an explicit form for these states, 
\begin{eqnarray}
&&\Phi _{s,z}^{\alpha ,\beta }\left( \eta \right) =\left[ \frac{\alpha }{%
\left| \alpha \right| \left( \alpha -\beta \right) }\right] ^{\frac{1}{2}%
}\left( \frac{\alpha ^{\ast }-\beta ^{\ast }}{\alpha -\beta }\right) ^{\frac{%
s}{2}}e^{Q_{4}}U_{s}\left( p_{2}\right) ,\;\;4\left| \alpha -\beta \right|
^{2}Q_{4}  \nonumber \\
&&\qquad \qquad =2\left( \alpha \beta ^{\ast }-\alpha ^{\ast }\beta \right)
\eta ^{2}+2\sqrt{2}\eta \left[ z\left( \alpha ^{\ast }-\beta ^{\ast }\right)
-z^{\ast }\left( \alpha -\beta \right) \right] +z^{\ast \;2}\left( \alpha
-\beta \right) ^{2}  \nonumber \\
&&\qquad \qquad \quad -z^{2}\left( \alpha ^{\ast }-\beta ^{\ast }\right)
^{2},\;\;2\left| \alpha -\beta \right| p_{2}=2\eta -\sqrt{2}z\left( \alpha
^{\ast }-\beta ^{\ast }\right) -\sqrt{2}z^{\ast }\left( \alpha -\beta
\right) \;.  \label{3.39}
\end{eqnarray}
The functions (\ref{3.39}) form a complete set for each fixed $z$ , 
\begin{equation}
\sum\limits_{s=0}^{\infty }\Phi _{s,z}^{\ast \;\alpha ,\beta }(\eta ^{\prime
})\Phi _{s,z}^{\alpha ,\beta }(\eta )=\delta (\eta ^{\prime }-\eta )\;,
\label{3.40}
\end{equation}
and for each fixed $s$ , 
\begin{equation}
\int \frac{d^{2}z}{\pi }\Phi _{s,z}^{\ast \;\alpha ,\beta }(\eta ^{\prime
})\Phi _{s,z}^{\alpha ,\beta }(\eta )=\delta (\eta -\eta ^{\prime }),\quad
d^{2}z=d%
\mathop{\rm Re}%
z\;d%
\mathop{\rm Im}%
z\;.  \label{3.41}
\end{equation}
The overlapping 
\begin{equation}
R_{s^{\prime },s}^{\alpha ^{\prime },\beta ^{\prime };\alpha ,\beta
}(z^{\prime },z)=\int_{-\infty }^{\infty }\Phi _{s^{\prime },z^{\prime
}}^{\ast \;\alpha ^{\prime },\beta ^{\prime }}(\eta )\Phi _{s,z}^{\alpha
,\beta }(\eta )\;d\eta \;,  \label{3.42}
\end{equation}
allows us to find mutual decompositions 
\begin{equation}
\Phi _{s,z}^{\alpha ,\beta }(\eta )=\sum\limits_{s^{\prime }=0}^{\infty
}R_{s^{\prime },s}^{\alpha ^{\prime },\beta ^{\prime };\alpha ,\beta
}(z^{\prime },z)\Phi _{s^{\prime },z^{\prime }}^{\alpha ^{\prime },\beta
^{\prime }}(\eta ),\;\Phi _{s,z}^{\alpha ,\beta }(\eta )=\int d^{2}z^{\prime
}\;R_{s^{\prime },s}^{\alpha ^{\prime },\beta ^{\prime };\alpha ,\beta
}(z^{\prime },z)\Phi _{s^{\prime },z^{\prime }}^{\alpha ^{\prime },\beta
^{\prime }}(\eta )\;.  \label{3.43}
\end{equation}
Unfortunately, the overlapping (\ref{3.42}) has a complicated form via a
finite sum of Hermit functions. In some particular cases this sum can be
simplified. For example, if $\alpha ^{\prime }=\alpha ,\;\beta ^{\prime
}=\beta ,$ then the overlapping does not depend on $\alpha ,\beta $ and has
the form 
\begin{equation}
R_{s^{\prime },s}^{\alpha ,\beta ;\alpha ,\beta }(z^{\prime
},z)=R_{s^{\prime },s}(z^{\prime },z)=\left( \frac{z-z^{\prime }}{z^{\ast
}-z^{\prime }{}^{\ast }}\right) ^{\frac{s\prime -s}{2}}\exp \left[ \frac{1}{2%
}\left( zz^{\prime }{}^{\;\ast }-z^{\ast }z^{\prime }\right) \right]
I_{s^{\prime },s}\left( |z-z^{\prime }|^{2}\right) .  \label{3.44}
\end{equation}
For $s=s^{\prime }=0$ \ we get 
\begin{eqnarray}
&&R_{0,0}^{\alpha ^{\prime },\beta ^{\prime };\alpha ,\beta }(z^{\prime },z)=%
\sqrt{\frac{\alpha \alpha ^{\prime }{}^{\ast }}{|\alpha \alpha ^{\prime }|}}%
\left( \alpha \alpha ^{\prime }{}^{\ast }-\beta \beta ^{\prime }{}^{\ast
}\right) ^{-\frac{1}{2}}\exp Q_{5}\,,  \nonumber \\
&&2Q_{5}=\frac{z^{2}(\alpha ^{\prime }{}^{\ast }\beta ^{\ast }-\alpha ^{\ast
}\beta ^{\prime }{}^{\ast })+(z^{\prime }{}^{\ast })^{2}(\alpha \beta
^{\prime }-\alpha ^{\prime }\beta )+2zz^{\prime }{}^{\ast }}{\alpha \alpha
^{\prime }{}^{\ast }-\beta \beta ^{\prime }{}^{\ast }}-z^{\prime }z^{\ast
}-zz^{\prime }{}^{\ast }-|z-z^{^{\prime }\ast }|^{2}\;.  \label{3.45}
\end{eqnarray}
The wave function $\Psi _{n,s,z}^{\alpha ,\beta }(x,y)$ has also a very
complicated form in the general case. However, in some particular cases it
can be simplified. For example, 
\begin{eqnarray}
&&\Psi _{n,s,z}^{1,0}(x,y)=\frac{(-1)^{n}}{\sqrt{\pi }}\left( \frac{x+iy-z}{%
x-iy-z^{\ast }}\right) ^{\frac{n-s}{2}}e^{M}I_{s,n}\left(
|x+iy-z|^{2}\right) \;,  \nonumber \\
&&M=z(x-iy)-z^{\ast }(x+iy)\;.  \label{3.46}
\end{eqnarray}
For $z=0$ we arrive at the set (\ref{3.19}).

For $s=0$ we get a compact form for a set of stationary squeezed coherent
states 
\begin{eqnarray}
&&\Psi _{n,0,z}^{\alpha ,\beta }(x,y)=\Psi _{n,z}^{\alpha ,\beta
}(x,y)=(-1)^{n}\frac{\pi ^{\frac{1}{4}}}{\sqrt{|\alpha |}}\left( \frac{\beta 
}{\alpha }\right) ^{\frac{n}{2}}U_{n}\left( \frac{p_{3}}{\sqrt{2\alpha \beta 
}}\right) \exp Q_{6}\;,  \nonumber \\
&&p_{3}=z-\alpha (x+iy)-\beta (x-iy)\;,  \label{3.47} \\
&&4\alpha \beta Q_{6}=(1+2|\beta |^{2})z^{2}-2\alpha \beta |z|^{2}+(z+p_{3}) 
\left[ \beta (x-iy)-\alpha (x+iy)\right] \,.\qquad  \nonumber
\end{eqnarray}
Additional simplifications are available for $\alpha =1,\,\beta =0,$ 
\begin{eqnarray}
&&\Psi _{n,z}^{1,0}(x,y)=\Psi _{n,z}(x,y)=\varphi _{n,z}(x,y)\exp \left( -%
\frac{1}{2}|z|^{2}\right) \;,  \nonumber \\
&&\varphi _{n,z}(x,y)=\frac{(x+iy-z)^{n}}{\sqrt{\pi \Gamma (n+1)}}\exp \left[
z(x-iy)-\frac{1}{2}\left( x^{2}+y^{2}\right) \right] \;.  \label{3.48}
\end{eqnarray}
Namely, these states where found in \cite{Jannu66}. However, the meaning of
the parameter $z$ was not clarified.

For arbitrary $\alpha ,\,\beta $, the functions $\Psi _{n,s,z}^{\alpha
,\beta }(x,y)$ obey, besides (\ref{3.7}), the following relations 
\begin{equation}
A_{2}^{\alpha ,\beta }\Psi _{n,s,z}^{\alpha ,\beta }=z\Psi _{n,s,z}^{\alpha
,\beta }+\sqrt{s}\Psi _{n,s-1,z}^{\alpha ,\beta }\;,\;\;A_{2}^{+\;\alpha
,\beta }\Psi _{n,s,z}^{\alpha ,\beta }=z^{\ast }\Psi _{n,s,z}^{\alpha ,\beta
}+\sqrt{s+1}\Psi _{n,s+1,z}^{\alpha ,\beta }\;.  \label{3.49}
\end{equation}
Taking into account that $a_{2}^{+}\varphi _{n,z}=\partial \varphi
_{n,z}/\partial z,$ we can construct a new set of stationary states by
successive differentiations, 
\begin{eqnarray}
&&\bar{\Psi}_{n,s,z}(x,y)=\frac{(-1)^{n}N}{\sqrt{\pi }}\exp \left[
i(n-s)\varphi +\frac{\rho -q}{2}\right] \left( \frac{q}{\rho }\right) ^{%
\frac{n-s}{2}}I_{s,n}(q)  \nonumber \\
&&\qquad \qquad \quad =\frac{(-1)^{n}N}{\sqrt{\pi }}\exp \left[ \frac{z}{2}%
(x+iy)\right] \left( \frac{x+iy-z}{x-iy}\right) ^{\frac{n-s}{2}}I_{s,n}(q)\;,
\nonumber \\
&&q=\rho -z\sqrt{\rho }e^{-i\varphi }=(x-iy)(x+iy-z)\;.  \label{3.51}
\end{eqnarray}
For $N=1$ the above set obeys (besides (\ref{3.7})) the relations 
\begin{equation}
a_{2}\bar{\Psi}_{n,s,z}=z\bar{\Psi}_{n,s,z}+\sqrt{s}\bar{\Psi}%
_{n,s-1,z}\;,\;\;a_{2}^{+}\bar{\Psi}_{n,s,z}=\frac{\partial }{\partial z}%
\bar{\Psi}_{n,s,z}=\sqrt{s+1}\bar{\Psi}_{n,s+1,z}\;.  \label{3.52}
\end{equation}
The set (\ref{3.19}) is a particular case of (\ref{3.51}), it corresponds to 
$z=0.$ The set (\ref{3.51}) is not orthogonal, 
\begin{eqnarray}
&&\left( \bar{\Psi}_{n^{\prime },s^{\prime },z^{\prime }},\bar{\Psi}%
_{n,s,z}\right) =N^{^{\prime }\ast }N\delta _{n,n^{\prime }}{\cal J}%
_{s,s^{\prime }}(z,z^{\prime })\;,  \nonumber \\
&&{\cal J}_{s,s^{\prime }}\left( z,z^{\prime }\right) =\sqrt{\frac{\Gamma
(s+1)}{\Gamma (s^{\prime }+1)}}z^{s^{\prime }-s}e^{zz^{\prime }{}^{\ast
}}L_{s}^{s^{\prime }-s}(-zz^{\prime }{}^{\ast }),\quad s\leq s^{\prime }\;, 
\nonumber \\
&&{\cal J}_{s,s^{\prime }}\left( z,z^{\prime }\right) =\sqrt{\frac{\Gamma
(s^{\prime }+1)}{\Gamma (s+1)}}(z^{\prime }{}^{\ast })^{s-s^{\prime
}}e^{zz^{\prime }{}^{\ast }}L_{s^{\prime }}^{s-s^{\prime }}(-zz^{\prime
}{}^{\ast }),\quad s^{\prime }\leq s\;.  \label{3.53}
\end{eqnarray}
The functions from the set (\ref{3.51}) are normalized to unity for $%
N=N_{s}(z)=\exp \left( -|z|^{2}/2\right) \left[ L_{s}\left( -|z|^{2}\right) %
\right] ^{-1/2}$. For $N=1$, the following mutual decompositions take place 
\begin{eqnarray}
&&\bar{\Psi}_{n,s+k,z^{\prime }}(x,y)=\sqrt{\frac{\Gamma (k+1)}{\Gamma
(s+k+1)}}\int \frac{d^{2}z}{\pi }z^{\ast s}e^{\left( z^{\prime }z^{\ast
}-|z|^{2}\right) }\bar{\Psi}_{n,k,z}(x,y),\quad \qquad  \nonumber \\
&&\bar{\Psi}_{n,s,z^{\prime }}(x,y)=\sum\limits_{k=0}^{\infty }\sqrt{\frac{%
\Gamma (k+s+1)}{\Gamma (s+1)}}\frac{(z^{\prime }-z)^{k}}{k!}\bar{\Psi}%
_{n,s+k,z}(x,y)\,.  \label{3.55}
\end{eqnarray}
That means, in particular, that (\ref{3.51}) is a complete set since the set
(\ref{3.19}) is complete.

Selecting different forms for the function $\Phi \left( \eta \right) ,$ we
can get other sets of stationary states for a charge in a uniform magnetic
field.

\subsection{Nonstationary states}

The most interesting nonstationary solutions of relativistic wave equations
for a charge in a uniform magnetic field are coherent states; for the first
time such solutions were presented in \cite
{BorKa74,BagBuG75,DodMaM75,BagBuG76}, see also \cite{BagGi90}. Below we
present a new family of nonstationary solutions, which includes the above
coherent states as a particular case.

Here we are going to use light-cone variables $u^{0}=x^{0}-x^{3},%
\;u^{3}=x^{0}+x^{3},$ and the corresponding momentum operators 
\begin{equation}
\widetilde{{\cal P}}_{0}=i\hbar \tilde{\partial}_{0}=\frac{1}{2}\left( {\cal %
P}_{0}-{\cal P}_{3}\right) ,\quad \widetilde{{\cal P}}_{3}=i\hbar \tilde{%
\partial}_{3}=\frac{1}{2}\left( {\cal P}_{0}+{\cal P}_{3}\right) \;, 
\nonumber
\end{equation}
where $\tilde{\partial}_{0}=\partial /\partial u^{0},\quad \tilde{\partial}%
_{3}=\partial /\partial u^{3}$. Then the Klein-Gordon operator can be
presented in the form 
\begin{equation}
{\cal K}=4\hbar ^{-2}\widetilde{{\cal P}}_{3}\widetilde{{\cal P}}%
_{0}-2\gamma {\cal N}-m^{\ast \;2}\;,  \label{4.3}
\end{equation}
whereas the Dirac equation reads (${\bf \Psi }$ is a Dirac bispinor) 
\begin{eqnarray}
&&4\hbar ^{-2}\widetilde{{\cal P}}_{3}\widetilde{{\cal P}}_{0}\Psi
_{(-)}=\left( 2\gamma {\cal N}_{D}+m^{\ast 2}\right) \Psi _{(-)}\;,\quad 2%
\widetilde{{\cal P}}_{3}\Psi _{(+)}=\left[ \left( {\bf \alpha }{\bf {\cal P}%
_{\perp }}\right) +\hbar \rho _{3}m\right] \Psi _{(-)}\;,  \nonumber \\
&&{\bf {\cal P}_{\perp }}=-\left( {\cal P}_{1},{\cal P}_{2},0\right)
\;,\quad {\bf \Psi }=\Psi _{\left( +\right) }+\Psi _{\left( -\right)
}\;,\quad \Psi _{\left( \pm \right) }=p_{\pm }{\bf \Psi },\quad 2p_{\pm
}=1\pm \alpha _{3}\;.  \label{4.4}
\end{eqnarray}
Here ${\bf \alpha }$ and $\rho _{3}$ are Dirac matrices \cite{BagGi90}, and $%
p_{\pm }$ projection operators.

In the case of the uniform magnetic field under consideration, the operators 
$\widetilde{{\cal P}}_{3},\widetilde{{\cal P}}_{0}$ are integrals of motion.
Thus, we will consider solutions that are eigenvectors of $\widetilde{{\cal P%
}}_{3},$ 
\begin{equation}
\widetilde{{\cal P}}_{3}\Psi =\hbar \frac{\lambda }{2}\Psi \,.  \label{4.5}
\end{equation}
The scalar wave function obeys (\ref{4.5}) and can be written as 
\begin{equation}
\Psi (x^{\mu })=N\exp \left( -i\frac{\lambda }{2}u^{3}-i\frac{m^{\ast \;2}}{%
2\lambda }u^{0}\right) \psi (u^{0},x,y)\;.  \label{4.6}
\end{equation}
It is easy to see that $\psi (u^{0},x,y)$ obeys a first order equation,
which can be treated as a Schr\"{o}dinger equation, 
\begin{equation}
i\partial _{0}\psi \left( u^{0},x,y\right) =\omega a_{1}^{+}a_{1}\psi \left(
u^{0},x,y\right) ,\quad \omega =\frac{\gamma }{\lambda }\;.  \label{4.7}
\end{equation}
Suppose Eq. (\ref{4.5}) holds, then $\Psi _{(-)}$ \ can be presented in the
form: 
\begin{equation}
\Psi _{(-)}(x^{\mu })=N\exp \left( -i\frac{\lambda }{2}u^{3}-i\frac{m^{\ast
\;2}}{2\lambda }u^{0}\right) W\left( 1-\alpha _{3}\right) C\psi
(u^{0},x,y)\;.  \label{4.8}
\end{equation}
Here $C$ is an arbitrary constant bispinor, and $W$ is a unitary matrix ($%
\varphi _{0}$ is a constant phase),

\begin{equation}
W=\cos \kappa -i\Sigma _{3}\sin \kappa ,\quad 2\kappa =\omega u^{0}+\varphi
_{0},\quad W^{+}W=I\;,  \label{4.9}
\end{equation}
and $\psi (u^{0},x,y)$ is a scalar function. The latter function obeys the
equation (\ref{4.7}). Then, the $\Psi _{(+)}$ projection can be found from (%
\ref{4.4}),\ $\Psi _{(+)}=\left( \hbar \lambda \right) ^{-1}\left[ \left( 
{\bf \alpha }{\cal P}_{\perp }\right) +\hbar m\rho _{3}\right] \Psi _{(-)}$.

Thus, both in the scalar and spinor cases we have to solve the same equation
(\ref{4.7}).

In the semi-momentum representation, the corresponding function $\tilde{\psi}%
(u^{0},\xi ,\eta )$ obeys the same equation (\ref{4.7}), where, however, one
has to use the expression (\ref{2.17}) for the operator ${\cal N}%
=a_{1}^{+}a_{1}.$ The relation between the functions $\tilde{\psi}(u^{0},\xi
,\eta )$ and $\psi (u^{0},\xi ,\eta )$ still has the form (\ref{2.20}).

Let us introduce the operators 
\begin{equation}
A_{1}^{f,g}=fa_{1}+ga_{1}^{+},\quad A_{1}^{+\;f,g}=f^{\ast
}a_{1}^{+}+g^{\ast }a_{1}\,,  \label{4.11}
\end{equation}
where the complex quantities $f$ and $g$ can depend on $u^{0}.$ These
operators are integrals of motion whenever $f,\,g$ obey the equations (by
dots above are denoted derivatives with respect to $u^{0}$) 
\begin{equation}
i\dot{f}+\omega f=0\,,\quad i\dot{g}-\omega g=0\;.  \label{4.12}
\end{equation}
It is \ easy to find 
\begin{equation}
f=f_{0}\exp \left( i\omega u^{0}\right) \,,\quad g=g_{0}\exp \left( -i\omega
u^{0}\right) \;,  \label{4.13}
\end{equation}
where $f_{0},\,g_{0}$ are some complex constants. Bearing in mind
considerations related to the operators (\ref{3.22}), we are going to
consider two nonequivalent cases only. The first one corresponds to $%
|f|^{2}=|g|^{2}$ or equivalently to $|f_{0}|^{2}=|g_{0}|^{2}.$ In this case
we can, in fact, only consider the Hermitian operator 
\begin{equation}
A_{1}^{\nu }=\nu a_{1}+\nu ^{\ast }a_{1}^{+},\quad \nu =\nu _{0}e^{i\omega
u^{0}},\quad \nu _{0}=\text{const\thinspace .}  \label{4.14}
\end{equation}
The second case corresponds to $|f|^{2}>|g|^{2},$ and here we can suppose
that 
\begin{equation}
|f|^{2}-|g|^{2}=|f_{0}|^{2}-|g_{0}|^{2}=1\;,  \label{4.15}
\end{equation}
without the loss of generality. In both cases the operators (\ref{4.11})
are, within constant complex factors, creation and annihilation operators.

Let us include operators (\ref{4.14}) and (\ref{3.23}) (they are integrals
of motion) into the complete set of operators. Then 
\begin{equation}
A_{1}^{\nu }\psi _{z_{1},z_{2}}^{\nu ,\mu }=z_{1}\psi _{z_{1},z_{2}}^{\nu
,\mu },\quad A_{2}^{\mu }\psi _{z_{1},z_{2}}^{\nu ,\mu }=z_{2}\psi
_{z_{1},z_{2}}^{\nu ,\mu },\quad z_{k}^{\ast }=z_{k},\quad k=1,2\;.
\label{4.16}
\end{equation}
In the semi-momentum representation we find 
\begin{equation}
\widetilde{\psi }\left( u^{0},\xi ,\eta \right) =\Phi _{z_{1}}^{\nu }\left(
\xi \right) \Phi _{z_{2}}^{\mu }\left( \eta \right) \;,  \label{4.17}
\end{equation}
where functions $\Phi _{z_{1}}^{\nu }$ are defined in (\ref{3.28}). The
corresponding coordinate representation reads 
\begin{eqnarray}
&&\psi _{z_{1},z_{2}}^{\nu ,\mu }(u^{0},x,y)=\left[ \frac{\mu \nu }{2\pi
^{2}|\mu ||\nu |(\mu \nu -\mu ^{\ast }\nu ^{\ast })}\right] ^{\frac{1}{2}%
}\exp \left[ \frac{Q_{6}}{4\left( \mu ^{\ast }\nu ^{\ast }-\mu \nu \right) }%
\right] \;,  \nonumber \\
&&Q_{6}=2(\mu +\mu ^{\ast })(\nu +\nu ^{\ast })x^{2}+2(\mu -\mu ^{\ast
})(\nu -\nu ^{\ast })y^{2}+4i(\mu \nu ^{\ast }-\mu ^{\ast }\nu )xy  \nonumber
\\
&&\qquad -4x\left[ z_{1}(\mu +\mu ^{\ast })+z_{2}(\nu +\nu ^{\ast })\right]
-4iy\left[ z_{1}(\mu -\mu ^{\ast })-z_{2}(\nu -\nu ^{\ast })\right] 
\nonumber \\
&&\qquad +\left( z_{1}\sqrt{\frac{\mu ^{\ast }}{\nu }}+z_{2}\sqrt{\frac{\nu 
}{\mu ^{\ast }}}\right) ^{2}+\left( z_{1}\sqrt{\frac{\mu }{\nu ^{\ast }}}%
+z_{2}\sqrt{\frac{\nu ^{\ast }}{\mu }}\right) ^{2}\;.\qquad  \label{4.18}
\end{eqnarray}
These solutions are orthogonal at any fixed $u^{0},\;\left( \psi
_{z_{1}^{\prime },z_{2}^{\prime }}^{\nu ,\mu },\psi _{z_{1},z_{2}}^{\nu ,\mu
}\right) =\delta \left( z_{1}-z_{1}^{\prime }\right) \delta \left(
z_{2}-z_{2}^{\prime }\right) $, and obey the completeness relation 
\begin{equation}
\int\limits_{-\infty }^{\infty }dz_{1}\int\limits_{-\infty }^{\infty
}dz_{2}\;\psi _{z_{1},z_{2}}^{\ast \;\nu ,\mu }(u^{0},x^{\prime },y^{\prime
})\psi _{z_{1},z_{2}}^{\nu ,\mu }(u^{0},x,y)=\delta \left( x-x^{\prime
}\right) \delta \left( y-y^{\prime }\right) \;.  \label{4.20}
\end{equation}

Consider now generalized squeezed coherent states, which can be constructed
by analogy with (\ref{3.38}) in the semi-momentum representation. We use
here the operators (\ref{4.11}) supposing that the relations (\ref{4.12}), (%
\ref{4.13}), (\ref{4.15}), and (\ref{3.24}) hold, 
\begin{equation}
\widetilde{\psi }_{n,s;z_{1},z_{2}}^{f,g;\alpha ,\beta }(u^{0},\xi ,\eta
)=\Phi _{n,z_{1}}^{f,g}\left( \xi \right) \Phi _{s,z_{2}}^{\alpha ,\beta
}\left( \eta \right) \,.  \label{4.21}
\end{equation}
The functions $\Phi _{n,z}^{a,b}\left( x\right) $ are defined in (\ref{3.39}%
). Thus, 
\begin{equation}
\psi _{n,s;z_{1},z_{2}}^{f,g;\alpha ,\beta }(u^{0},x,y)=\frac{\left(
A_{1}^{+\;f,g}-z_{1}^{\ast }\right) ^{n}\left( A_{2}^{+\;\alpha ,\beta
}-z_{2}^{\ast }\right) ^{s}}{\sqrt{\Gamma (n+1)\Gamma (s+1)}}\psi
_{z_{1},z_{2}}^{f,g;\alpha ,\beta }(u^{0},x,y),\,\,\psi
_{z_{1},z_{2}}^{f,g;\alpha ,\beta }=\psi _{0,0;z_{1},z_{2}}^{f,g;\alpha
,\beta }\,.  \label{4.21a}
\end{equation}
The solutions (\ref{4.21a}) obey the relations 
\begin{eqnarray}
&&\left( A_{1}^{f,g}-z_{1}\right) \psi _{n,s;z_{1},z_{2}}^{f,g;\alpha ,\beta
}=\sqrt{n}\psi _{n-1,s;z_{1},z_{2}}^{f,g;\alpha ,\beta }\,,\,\left(
A_{1}^{+\;f,g}-z_{1}^{\ast }\right) \psi _{n,s;z_{1},z_{2}}^{f,g;\alpha
,\beta }=\sqrt{n+1}\psi _{n+1,s;z_{1},z_{2}}^{f,g;\alpha ,\beta }\;, 
\nonumber \\
&&\left( A_{2}^{\alpha ,\beta }-z_{2}\right) \psi
_{n,s;z_{1},z_{2}}^{f,g;\alpha ,\beta }=\sqrt{s}\psi
_{n,s-1;z_{1},z_{2}\,}^{f,g;\alpha ,\beta },\,\left( A_{2}^{+\;\alpha ,\beta
}-z_{2}^{\ast }\right) \psi _{n,s;z_{1},z_{2}}^{f,g;\alpha ,\beta }=\sqrt{s+1%
}\psi _{n,s+1;z_{1},z_{2}}^{f,g;\alpha ,\beta }\,.  \label{4.22}
\end{eqnarray}
Eq. (\ref{4.21a}) describes the most general form of relativistic wave
equation solutions in a constant uniform magnetic field. \ All the formerly
known solutions can be obtained from this equation by a particular choice of
parameters. For instance, by selecting $f_{0}=\alpha =1,\;g=\beta
=0,\;z_{1}=0,$\ $z_{2}=z$ with $z=0$ we get \ the states (\ref{3.19}), on
the other hand, if one puts $s=0,\;z\neq 0$, then one gets the states (\ref
{3.48}). For $n=s=0,\;f_{0}=\alpha =1,\;g=\beta =0$, we get coherent states 
\cite{BorKa74,BagBuG75,DodMaM75,BagBuG76}.

In the general case, an explicit coordinate representation for the solutions
(\ref{4.21a}) looks complicated enough. However, some particular cases admit
essential simplifications. For example, suppose $f_{0}=\alpha =1,\;g=\beta
=0 $, then 
\begin{eqnarray}
&&\Psi _{n,s;z_{1},z_{2}}^{1,0;1,0}(u^{0},x,y)=\frac{(-1)^{n}}{\sqrt{\pi }}%
\left( \frac{x+iy-\bar{z}_{1}^{\ast }-z_{2}}{x-iy-\bar{z}_{1}-z_{2}^{\ast }}%
\right) ^{\frac{n-s}{2}}e^{M_{1}}I_{s,n}(p_{4})\;,  \nonumber \\
&&2M_{1}=\left( \bar{z}_{1}-z_{2}^{\ast }\right) \left( x+iy\right) -\left( 
\bar{z}_{1}^{\ast }-z_{2}\right) \left( x-iy\right) +\bar{z}_{1}^{\ast
}z_{2}-\bar{z}_{1}z_{2}^{\ast }-2in\omega u^{0}\;,  \nonumber \\
&&p_{4}=|x+iy-\bar{z}_{1}^{\ast }-z_{2}|^{2}\,,\quad \bar{z}_{1}=z_{1}\exp
(-i\omega u^{0})\;,  \label{4.23}
\end{eqnarray}
For $n=s=0$, we get the coordinate representation for the squeezed coherent
states in the form 
\begin{eqnarray}
&&\Psi _{z_{1},z_{2}}^{f;g;\alpha ,\beta }(u^{0},x,y)=\left[ \frac{\alpha f}{%
(\alpha f-\beta g)\pi |\alpha ||f|}\right] ^{\frac{1}{2}}\exp Q_{7}\;, 
\nonumber \\
&&Q_{7}=-\frac{1}{2}\left( |z_{1}|^{2}+|z_{2}|^{2}\right) +\frac{q}{2(\alpha
f-\beta g)}\quad ,  \nonumber \\
&&q=-(\alpha +\beta )(f+g)x^{2}-(\alpha -\beta )(f-g)y^{2}+2i(\beta f-\alpha
g)xy  \nonumber \\
&&\qquad +2x\left[ (\alpha +\beta )z_{1}+(f+g)z_{2}\right] +2iy\left[
(\alpha -\beta )z_{1}-(f-g)z_{2}\right]  \nonumber \\
&&\qquad +(\alpha g^{\ast }-\beta f^{\ast })z_{1}^{2}+(\beta ^{\ast
}f-\alpha ^{\ast }g)z_{2}^{2}-2z_{1}z_{2}\;.\qquad  \label{4.24}
\end{eqnarray}
Solutions from \cite{BorKa74,BagBuG75,DodMaM75,BagBuG76} are particular
cases of (\ref{4.24}) for $f_{0}=\alpha =1,\;g=\beta =0.$

Calculating mean values in the states (\ref{4.21a}), we get\footnote{%
One can obtain the same results using spinor wave functions for the
calculations.} 
\begin{equation}
\overline{{\cal P}}_{1}=i\hbar \sqrt{\frac{\gamma }{2}}\left[ \left( f^{\ast
}+g^{\ast }\right) z_{1}-\left( f+g\right) z_{1}^{\ast }\right] \;,\quad 
\overline{{\cal P}}_{2}=-\hbar \sqrt{\frac{\gamma }{2}}\left[ \left( f^{\ast
}-g^{\ast }\right) z_{1}+\left( f-g\right) z_{1}^{\ast }\right] \;.
\label{4.25}
\end{equation}
Here we have taken into account the relations (\ref{2.7}), (\ref{3.25}), (%
\ref{4.22}), and the orthogonality of the states with respect to the indices 
$n,s.$ Remember now that in classical theory the corresponding momenta $%
{\cal P}_{1}^{cl},{\cal P}_{2}^{cl}$ have the following parametric
representation (with $u^{0}$ being the evolution parameter, $R$ radius of
the classical orbit, and $\kappa $ is given by (\ref{4.9})) 
\begin{equation}
{\cal P}_{1}^{cl}=\hbar \gamma R\sin 2\kappa ,\quad {\cal P}_{2}^{cl}=-\hbar
\gamma R\cos 2\kappa \;.  \label{4.26}
\end{equation}
It is easy to see that (\ref{4.25}) coincides with (\ref{4.26}) for $%
z_{1}=\left( \gamma /2\right) ^{1/2}R\left( f_{0}e^{-i\varphi
_{0}}+g_{0}e^{i\varphi _{0}}\right) $. Calculating mean values of the
coordinates $\overline{x^{1}},\overline{\,x^{2}},$ we find that they evolve
as the corresponding classical quantities $x^{1\,cl},\,x^{2\;cl}$ ( $%
x_{\left( 0\right) }^{1},x_{\left( 0\right) }^{2}$ are coordinates of the
orbit center) 
\begin{equation}
x^{1\;cl}=R\cos \kappa +x_{\left( 0\right) }^{1},\quad x^{2\;cl}=R\sin
\kappa +x_{\left( 0\right) }^{2}\;,  \label{4.28}
\end{equation}
for $z_{2}=\left( \gamma /2\right) ^{1/2}\left[ (\alpha +\beta
)x_{(0)}^{1}+i(\alpha -\beta )x_{(0)}^{2}\right] $.

Thus, mean-value trajectories in the plane $x^{1},x^{2}$ do not depend on
quantum numbers $n,s$. These trajectories have classical forms under a
proper choice of $z_{1},z_{2}$ .

Calculating quadratic fluctuations in the states (\ref{4.21a}), we get 
\begin{eqnarray}
&&2\overline{\left( \Delta {\cal P}_{1}\right) ^{2}}=\hbar ^{2}\gamma
|f+g|^{2}(2n+1),\quad 2\overline{\left( \Delta {\cal P}_{2}\right) ^{2}}%
=\hbar ^{2}\gamma \left| f-g\right| ^{2}(2n+1)\;,  \nonumber \\
&&2\gamma \overline{\left( \Delta x^{1}\right) ^{2}}=|f-g|^{2}(2n+1)+|\alpha
-\beta |^{2}(2s+1)\;,  \nonumber \\
&&2\gamma \overline{\left( \Delta x^{2}\right) ^{2}}=|f+g|^{2}(2n+1)+|\alpha
+\beta |^{2}(2s+1)\;,  \nonumber \\
&&\sigma _{1}\;=-\sigma _{2}=i\left( fg^{\ast }-gf^{\ast }\right) (2n+1)\;, 
\nonumber \\
&&\sigma _{k}=\overline{\left( \Delta x^{k}\right) \left( \Delta
P_{k}\right) +\left( \Delta P_{k}\right) \left( \Delta x^{k}\right) },\quad
k=1,2\;.  \label{4.30}
\end{eqnarray}
They do not depend on $z_{1},z_{2},$ but do depend on quantum numbers $n,s$
and on parameters $f_{0},g_{0},\alpha ,\beta .$ The relations (\ref{4.30})
imply the generalized Heisenberg inequalities 
\begin{eqnarray}
&&4{\cal J}_{1}=\hbar ^{2}(2n+1)\left[ (2n+1)+\left( 2s+1\right) |(\alpha
-\beta )(f+g)|^{2}\right] \geq \hbar ^{2}\;,  \nonumber \\
&&4{\cal J}_{2}=\hbar ^{2}(2n+1)\left[ (2n+1)+\left( 2s+1\right) |(\alpha
+\beta )(f-g)|^{2}\right] \geq \hbar ^{2}\;.  \nonumber \\
&&{\cal J}_{k}=\overline{\left( \Delta x^{k}\right) ^{2}}\;\overline{\left(
\Delta {\cal P}_{k}\right) ^{2}}-\frac{1}{4}\sigma _{k}^{2},\quad k=1,2\;.
\label{4.31}
\end{eqnarray}
One can fix any given$\overline{\;\left( \Delta x^{k}\right) ^{2}}\;$or$\ \ 
\overline{\left( \Delta {\cal P}_{k}\right) ^{2}}$ in a given ''instant'' $%
u^{0}$ by means of a choice of parameters $f_{0},g_{0},\alpha ,\beta $ .
Then they evolve with ''time'' $u^{0}$ according Eqs. (\ref{4.30}).

Below we present another type of nonstationary states, which are quite
different from the above generalized coherent states. Recall that the
problem was reduced to solving the equation (\ref{4.7}) under the condition (%
\ref{4.5}). All the integrals of motion for such an equation can be
constructed as functional combination of the operators 
\begin{equation}
fa_{1},\quad ga_{1}^{+},\quad a_{2},\quad a_{2}^{+},  \label{4.32}
\end{equation}
whenever $f,g$ obey the relations (\ref{4.12}), (\ref{4.13}). Constructing
integrals of motion that are linear combinations of these operators, we get
coherent states. Any linear combinations of the operators (\ref{4.32}) do
not commute with the operator $L_{z}$ \ (\ref{2.11}) or $J_{z}$ (\ref{2.4}).
Thus, coherent states with definite values of these quantities cannot be
constructed. The generalized squeezed coherent states (\ref{3.38}) and (\ref
{4.21}) are eigenvectors of the operators ${\cal N}_{1},{\cal N}_{2}$ (that
follows from (\ref{4.22}). The latter operators are integrals of motion and
are quadratic in creation and annihilation operators,

\begin{equation}
{\cal N}_{1}=\left( A_{1}^{+\;f,g}-z_{1}^{\ast }\right) \left(
A_{1}^{f,g}-z_{1}\right) ,\quad {\cal N}_{2}=\left( A_{2}^{+\;\alpha ,\beta
}-z_{2}^{\ast }\right) \left( A_{2}^{\alpha ,\beta }-z_{2}\right) .
\label{4.33}
\end{equation}
The operators $A_{1}^{f,g}$ are defined in (\ref{4.11}), and $A_{2}^{\alpha
,\beta }$ are defined in (\ref{3.24}). The operators (\ref{4.33}) do not
commute with $L_{z\ },$ $J_{z}$ as well.

One can see that besides the operators $a_{1}a_{2},\;a_{1}^{+}a_{2}^{+}$ ,
the only one quadratic combination that commutes with $L_{z\ },$ $J_{z}$ is 
\begin{equation}
\bar{A}=fa_{1}a_{2}+ga_{1}^{+}a_{2}^{+}\,.  \label{4.34}
\end{equation}
It is known \cite{BagGiK75} that eigenvectors for such an operator can be
normalized only for $|f|>|g|,$ or for $|f|=|g|.$ In the first case the
eigenvectors\ have a finite norm, and in the second case they can be
normalized to a $\delta -$function. Let us consider the case $|f|\geq |g|$ \
only. Here the operator (\ref{4.34}) differs from 
\begin{equation}
A^{p}=e^{i\kappa }\left( a_{1}a_{2}-\bar{p}^{2}a_{1}^{+}a_{2}^{+}\right)
,\,\,\bar{p}=pe^{-i\kappa },\,-1\leq p\leq 1,\;\kappa =\omega u^{0}+\kappa
_{0},\,\kappa _{0}={\rm const}  \label{4.35}
\end{equation}
by a complex factor only. Thus, it is enough to consider the latter operator
only. Let us demand that functions $\psi \left( u^{0},\rho ,\varphi \right) $
be solutions of the equation (\ref{4.7}), and, at the same time,
eigenvectors of the operators $A^{p},\quad L_{z},$%
\begin{equation}
A^{p}\psi _{q,l}^{p}=-q\psi _{q,l}^{p},\quad L_{z}\psi _{q,l}^{p}=\hbar
l\psi _{q,l}^{p},\quad l=0,\pm 1,\pm 2,\ldots \;.  \label{4.36}
\end{equation}
Such solutions can be constructed in terms of the Laguerre functions $%
I_{n,m}\left( x\right) $ with noninteger indices, 
\begin{eqnarray}
&&\psi _{q,l}^{p}\left( u^{0},\rho ,\varphi \right) =N\exp \left( il\varphi
-\Gamma \right) \left( 1+\bar{p}\right) ^{-\alpha }\left( 1-\bar{p}\right)
^{-\beta }I_{\left| l\right| +s,s}\left( x\right) \;,  \nonumber \\
&&\Gamma =i\frac{l\omega u^{0}}{2}+\frac{1+\bar{p}^{2}}{2\left( 1-\bar{p}%
^{2}\right) }\rho ,\quad \alpha =\frac{p-q}{2p},\quad \beta =\frac{p+q}{2p}%
\;,  \nonumber \\
&&s=\frac{q}{2p}-\frac{\left| l\right| +1}{2},\quad x=\frac{2\bar{p}\rho }{1-%
\bar{p}^{2}}\;,  \nonumber \\
&&I_{n,m}\left( x\right) =\sqrt{\frac{\Gamma \left( 1+n\right) }{\Gamma
\left( 1+m\right) }}\frac{\exp \left( -\frac{x}{2}\right) }{\Gamma \left(
1+n-m\right) }x^{\frac{n-m}{2}}\Phi \left( -m,n-m+1;x\right) \;.\qquad
\label{4.37}
\end{eqnarray}
Here $\Phi \left( \alpha ,\beta ;x\right) $ is the degenerate hypergeometric
function (see \cite{GraRy94}, 9.210). For $p^{2}=1$, the operator (\ref{4.35}%
) is anti-Hermitian and $q$ is imaginary, $%
\mathop{\rm Re}%
\,q=0.$ For $p=0$, solutions have a very simple form 
\begin{equation}
\psi _{q,l}^{0}\left( u^{0},\rho ,\varphi \right) =N_{0}\exp \left(
il\varphi +\bar{q}-\Gamma _{0}\right) J_{\left| l\right| }\left( 2\sqrt{\bar{%
q}\rho }\right) \;,\quad \Gamma _{0}=\frac{i}{2}l\omega u^{0}+\frac{\rho }{2}%
,\quad \bar{q}e^{-ik}\;,  \label{4.38}
\end{equation}
where $J_{\nu }(x)$ is the Bessel function (see \cite{GraRy94}, 8.402). The
functions (\ref{4.38}) can be obtained from (\ref{4.37}) as a limit $%
p\rightarrow 0$, as can be seen with the help of the property 
\begin{equation}
\lim_{r\rightarrow \infty }I_{r+\alpha ,r+\beta }\left( \frac{x^{2}}{4r}%
\right) =J_{\alpha -\beta }\left( x\right) \;.  \label{4.39}
\end{equation}
The functions (\ref{4.37}) and (\ref{4.38}) are orthogonal only with respect
to quantum numbers $l$, 
\begin{eqnarray}
&&\left( \psi _{q^{\prime },l^{\prime }}^{p},\psi _{q,l}^{p}\right) =\delta
_{l,l^{\prime }}QF\left( -s,-s^{\prime \ast };1+\left| l\right| ;y\right)
,\quad y=\left( \frac{2p}{1+p^{2}}\right) ^{2}\;,  \nonumber \\
&&Q=\left[ \frac{\Gamma \left( 1+\left| l\right| +s\right) \Gamma \left(
1+\left| l\right| +s^{^{\prime }\ast }\right) }{p^{2}\Gamma \left(
1+s\right) \Gamma \left( 1+s^{^{\prime }\ast }\right) }\right] ^{\frac{1}{2}}%
\frac{\pi NN^{^{\prime }\ast }}{\Gamma \left( 1+\left| l\right| \right) }y^{%
\frac{1+\left| l\right| }{2}}\left( 1-y\right) ^{-\frac{q+q^{\ast }}{4p}}\;,
\nonumber \\
&&\left( \psi _{q^{\prime },l^{\prime }}^{0},\psi _{q,l}^{0}\right) =\delta
_{l,l^{\prime }}2\pi N_{0}N_{0}^{^{\prime }\ast }I_{\left| l\right| }\left( 2%
\sqrt{qq^{^{\prime }\ast }}\right) \;.  \label{4.40}
\end{eqnarray}
Here $F\left( \alpha ,\beta ;\gamma ;x\right) $ is the hypergeometric
function (see \cite{GraRy94}, 9.100), and $I_{\alpha }(x)$ is the Bessel
function of imaginary argument (see \cite{GraRy94}, 8.404). Calculating (\ref
{4.40}), we have used the integral table (see \cite{GraRy94}, 6.633.2;
7.622.1).

The states (\ref{4.37}) are not coherent states, however, they are, in a
sense, close to such states. Indeed, let us consider the equations (\ref{2.7}%
) on classical trajectories. Then we get a classical relation 
\begin{equation}
\rho =\rho \left( u^{0}\right) =\sqrt{L_{z}^{2}\hbar ^{-2}+4\left|
a_{1}a_{2}\right| ^{2}}-a_{1}a_{2}-a_{1}^{+}a_{2}^{+}\;.  \label{4.41}
\end{equation}
For $p=0,$ it follows from (\ref{4.36}) that $a_{1}a_{2}=-\bar{q},\quad
L_{z}=\hbar l.$ Thus, we can rewrite (\ref{4.41}) in the form 
\begin{equation}
\rho \left( u^{0}\right) =\rho _{0}^{cl}+\bar{q}+\bar{q}^{\ast },\quad \rho
_{0}^{cl}=\sqrt{l^{2}+4\left| q\right| ^{2}}\;.  \label{4.43}
\end{equation}
Calculating the mean value $\bar{\rho}$ by means of the functions (\ref{4.38}%
), we find 
\begin{equation}
\bar{\rho}=\rho _{0}+\bar{q}+\bar{q}^{\ast },\quad \rho _{0}=\left| l\right|
-1-2\left| q\right| \frac{I_{\left| l\right| -1}\left( 2\left| q\right|
\right) }{I_{\left| l\right| }\left( 2\left| q\right| \right) }\;.
\label{4.44}
\end{equation}
Thus, the time dependence of $\bar{\rho}$ is classical. The only constant
which can differ from its classical value is $\rho _{0}$ .

\section{Exact solutions of relativistic wave equations in longitudinal
fields}

\subsection{Definition of fields}

Consider here longitudinal electromagnetic fields, which have the form

\begin{equation}
{\bf E}={\bf n}E,\quad {\bf H}={\bf n}H\,.  \label{4.45}
\end{equation}
Here ${\bf n}$ is a unit vector, ${\bf n}^{2}=1$. Suppose ${\bf n}$ is
directed along $x^{3}$ axis. Then, the fields (\ref{4.45}) obey the free
Maxwell equations whenever

\[
E=E\left( x^{0},x^{3}\right) ,\quad H=H\left( x^{1},x^{2}\right) , 
\]
where $E\left( x^{0},x^{3}\right) ,\;H\left( x^{1},x^{2}\right) $ are
arbitrary functions of the indicated arguments. Thus, the fields under
consideration can be represented by potentials of the form

\begin{eqnarray}
&&A_{0}=A_{0}\left( x^{0},x^{3}\right) ,\;A_{1}=A_{1}\left(
x^{1},x^{2}\right) ,\;A_{3}=A_{3}\left( x^{0},x^{3}\right)
\,,\;A_{2}=A_{2}\left( x^{1},x^{2}\right) ,  \nonumber \\
\quad \; &&E=\partial _{0}A_{3}-\partial _{3}A_{0},\;H=\partial
_{2}A_{1}-\partial _{1}A_{2}\,.  \label{4.46}
\end{eqnarray}
Thus, the operators (\ref{2.7}) do not depend on the electric field (on $%
A_{0},\,A_{3}$). Therefore, imposing restrictions only on the magnetic
field, we can maintain the relations (\ref{2.7}-\ref{2.14}). For a uniform
magnetic field (\ref{2.1}), the commutation relations (\ref{2.10}) are still
valid and we use the semi-momentum representation, where these operators act
on different variables (see (\ref{2.13}-\ref{2.15}), (\ref{2.20})).

Lorentz equations have the following form

\begin{eqnarray}
m\ddot{x}^{0}+E\dot{x}^{3} &=&0,\quad m\ddot{x}^{3}+E\dot{x}^{0}=0; 
\nonumber \\
m\ddot{x}^{1}+H\dot{x}^{2} &=&0,\quad m\ddot{x}^{2}+H\dot{x}^{1}=0,\;\dot{x}%
^{\mu }\dot{x}_{\mu }=1\,,  \label{4.46a}
\end{eqnarray}
which implies the following first integrals of motion,

\begin{equation}
m^{2}\left( \left( \dot{x}_{1}\right) ^{2}+\left( \dot{x}_{2}\right)
^{2}\right) =k_{1}^{2},\quad m^{2}\left( \left( \dot{x}_{0}\right)
^{2}-\left( \dot{x}_{3}\right) ^{2}\right) =m^{2}+\left( k_{1}\right) ^{2},
\label{4.47}
\end{equation}
where $k_{1}$ is an integration constant.

\subsection{Klein-Gordon equation}

Consider the Klein-Gordon equation in the fields under consideration.
Representing the wave function as 
\begin{equation}
\Psi =\varphi \left( x^{0},x^{3}\right) \psi \left( x^{1},x^{2}\right) ,
\label{5.1}
\end{equation}
we find 
\begin{equation}
\left( {\cal P}_{1}^{2}+{\cal P}_{2}^{2}-k_{1}^{2}\right) \psi \left(
x^{1},x^{2}\right) =0,\;\left( {\cal P}_{0}^{2}-{\cal P}%
_{3}^{2}-m^{2}-k_{1}^{2}\right) \varphi \left( x^{0},x^{3}\right) =0\,.
\label{5.2}
\end{equation}
Using the variables $x,\;y$, $\eta $, $\xi $ defined in (\ref{2.5}) and (\ref
{2.15}), we can rewrite the equation for $\psi \left( x^{1},x^{2}\right) $
in the following form

\begin{equation}
\left( \xi ^{2}-\partial _{\xi }^{2}-k_{1}^{\prime 2}\right) \psi \left(
x^{1},x^{2}\right) =0,\quad k_{1}^{\prime 2}=\frac{k_{1}^{2}}{\hbar
^{2}\gamma }\,.  \label{5.5}
\end{equation}
The operator $\left( \xi ^{2}-\partial _{\xi }^{2}-k_{1}^{\prime 2}\right) $
does not depend on $\eta $ . Thus, it is convenient to go over to the
semi-momentum representation,

\[
\psi \left( x,y\right) =\frac{1}{\sqrt{2\pi }}\int\limits_{-\infty }^{\infty
}e^{iky}\tilde{\psi}\left( x,k\right) \;dk. 
\]
Substituting the integration over $k$ by an integration over $\eta $, we
obtain

\begin{equation}
\psi \left( x,y\right) =\frac{e^{ixy}}{\sqrt{\pi }}\int\limits_{-\infty
}^{\infty }e^{-i\sqrt{2y}\eta }\tilde{\psi}\left( \xi ,\eta \right) \;d\eta
,\quad \xi =\sqrt{2}x-\eta .  \label{5.6}
\end{equation}
The function $\psi $ is an eigenvector of the operator ${\cal N}$ (\ref{2.17}%
). The latter operator commutes with the operators from Eqs. (\ref{5.2}). In
the semi-momentum representation we can write

\begin{equation}
{\cal N}\tilde{\psi}_{n}=n\tilde{\psi}_{n}\Rightarrow \tilde{\psi}_{n}\left(
\xi ,\eta \right) =U_{n}\left( \xi \right) \Phi \left( \eta \right) ,
\label{5.7}
\end{equation}
where $\Phi \left( \eta \right) $ is an arbitrary function of $\eta $. It
follows from (\ref{2.11}) and the first Eq. (\ref{5.2}) that

\[
{\cal P}_{1}^{2}+{\cal P}_{2}^{2}=2\hbar ^{2}\gamma {\cal N}+\hbar
^{2}\gamma ,\;k_{1}^{2}=2\hbar ^{2}\gamma n+\hbar ^{2}\gamma \;. 
\]
Solution of the last equation (\ref{5.2}) can be found for fields that admit
separation of the variables $x^{0},\,x^{3}$. For example, let us choose the
potentials in the form: $\left| e\right| A_{0}=A\left( x^{3}\right)
,\,\,A_{3}=0,\;\left| e\right| E=-\partial _{3}A$. In this case, stationary
solutions of Eq. (\ref{5.2}) read 
\begin{equation}
\varphi \left( x^{0},x^{3}\right) =e^{-ik_{0}x^{0}}\chi \left( x^{3}\right)
,\quad \chi ^{\prime \prime }+R\chi =0,\quad R\left( x^{3}\right) =\left(
k_{0}+A\right) ^{2}-m^{2}-k_{1}^{2}\,.  \label{5.8}
\end{equation}
Thus, the functions (\ref{5.1}) being written in the semi-momentum
representation take the form

\begin{equation}
\Psi _{n}=e^{-ik_{0}x^{0}}\chi \left( x^{3}\right) U_{n}\left( \xi \right)
\Phi \left( \eta \right) \,.  \label{5.9}
\end{equation}
The equation (\ref{5.8}) for $\chi $ can be solved exactly for the following
choices of the function $A\left( x^{3}\right) $:

\begin{eqnarray*}
A\left( x^{3}\right) &=&\alpha x,\quad A\left( x^{3}\right) =\alpha \exp
\left( \beta x^{3}\right) ,\quad A\left( x^{3}\right) =\frac{\alpha }{x^{3}},
\\
A\left( x^{3}\right) &=&\alpha \tanh \left( \beta x^{3}\right) ,\quad
A\left( x^{3}\right) =\alpha \tan \left( \beta x^{3}\right) ,\quad A\left(
x^{3}\right) =\beta \coth \left( \beta x^{3}\right) \,.
\end{eqnarray*}
The corresponding exact solutions are presented in \cite{BagGi90}. Demanding
that the Klein-Gordon function be an eigenvector of the operator $L_{z}$ (%
\ref{2.19}), we get an equation for the function $\Phi \left( \eta \right) $
from (\ref{5.9}),

\begin{equation}
a_{2}^{+}a_{2}\Phi \left( \eta \right) =s\Phi \left( \eta \right) \,.
\label{5.9a}
\end{equation}
Thus, $\Phi \left( \eta \right) =U_{s}\left( \eta \right) ,$ and $L_{Z}\Psi
_{ns}=\hbar \left( n-s\right) \Psi _{ns}\,.$ Keeping this in mind and doing
the integral (\ref{5.6}), we obtain

\[
\Psi _{n,s}=e^{-ik_{0}x^{0}}\chi \left( x^{3}\right) \frac{\left( -1\right)
^{n}}{\sqrt{2\pi }}\left( \frac{x+iy}{x-iy}\right) ^{\frac{n-s}{2}%
}I_{n,s}\left( x^{2}+y^{2}\right) \,, 
\]
where $I_{n,s}$ are the Laguerre functions. Solutions of the equation (\ref
{5.9a}) have been analyzed above.

\subsection{Dirac equation}

Let us present the Dirac wave function in the form

\begin{eqnarray*}
&&\Psi ={\bf M}\left( 
\begin{array}{c}
\psi _{1} \\ 
\psi _{-1}
\end{array}
\right) \varphi \left( x^{0},x^{3}\right) v, \\
&&{\bf M}=\left( 
\begin{array}{cc}
k_{1}\left( m+{\cal P}_{0}+{\cal P}_{3}-ik_{1}\sigma _{2}\right) & \left( 
{\cal P}_{1}-i{\cal P}_{2}\right) \left( m+{\cal P}_{0}+{\cal P}%
_{3}-ik_{1}\sigma _{2}\right) \\ 
\left( {\cal P}_{1}+i{\cal P}_{2}\right) \left[ \left( m-{\cal P}_{0}-{\cal P%
}_{3}\right) \sigma _{3}-k_{1}\sigma _{1}\right] & k_{1}\left[ \left( m-%
{\cal P}_{0}-{\cal P}_{3}\right) \sigma _{3}-k_{1}\sigma _{1}\right]
\end{array}
\right) ,
\end{eqnarray*}
where $\nu $ is an arbitrary spinor, which can be fixed by supplementary
conditions. Then the following equations take place

\begin{eqnarray}
&&\left( {\cal P}_{0}^{2}-{\cal P}_{3}^{2}-m^{2}-k^{\prime }\right) \varphi
\left( x^{0},x^{3}\right) =0,\quad k^{\prime }=k_{1}^{2}+ieE,  \label{5.10}
\\
&&\left( {\cal P}_{1}+i{\cal P}_{2}\right) \psi _{1}\left(
x^{1},x^{2}\right) =\hbar k_{1}\psi _{-1}\,\left( x^{1},x^{2}\right) ,
\label{5.11} \\
&&\left( {\cal P}_{1}-i{\cal P}_{2}\right) \psi _{-1}\left(
x^{1},x^{2}\right) =\hbar k_{1}\psi _{1}\left( x^{1},x^{2}\right) .
\label{5.12}
\end{eqnarray}
As a consequence of Eqs. (\ref{5.11}) and (\ref{5.12}), we get

\[
a_{1}\psi _{-1}=-i\sqrt{n}\psi _{1}\,,\;a_{1}^{+}\psi _{1}=i\sqrt{n}\psi
_{-1},\;\;k_{1}^{2}=2\gamma n. 
\]
Thus, we see that $\psi _{1}=\psi _{n-1}\,,\;\psi _{-1}=-i\psi _{n},$ and
the problem is reduced to solving the equation (\ref{5.10}). The latter
coincides with the second equation (\ref{5.2}).

Considering, for example, the constant and uniform magnetic field (\ref{2.1}%
) together with the electric field described by potentials $\left| e\right|
A_{0}=A\left( x^{3}\right) ,\quad A_{3}=0\Rightarrow \left| e\right|
E=-\partial _{3}A$, we get

\[
\varphi \left( x^{0},x^{3}\right) =\exp \left( -ik_{0}x^{0}\right) \chi
\left( x^{3}\right) ,\quad \chi ^{\prime \prime }+\left( i\partial
_{3}+k_{0}+{\cal A}\right) \chi =0\,. 
\]
All possible solutions of the latter equation for the function $\chi $ are
presented in \cite{BagGi90}.

\section*{Peculiarities of integration of linear differential equations with
noncommutative symmetries}

Here we are going to return to the above results from a point of view of
general theory of differential equation. Recall that we succeeded to find
explicitly the transformation (\ref{2.13})-(\ref{2.16}) which has reduced
effectively the number of the variables in the initial equations. In fact,
that was the main starting point for all further constructions. However, one
can see that this ''reduction'' of variables is a particular example of a
general situation, which is described briefly below.

Consider first the case of an integrable classical $2N$-dimensional
Hamiltonian system with the Hamiltonian $H$. Suppose this system has $N$
independent integrals of motion that are in involution. It is well known
that in such a case the variables of the type action-angle $(J,\varphi )$
are available, and the Hamiltonian depends on the action variables only, $%
H=H(J)$. Let us suppose that for such a system exists one more independent
integral of motion $Y$. Since $Y$ is independent, it cannot commute with the
former integrals, and, therefore, $Y$ must depend on the angle variables.
One can demonstrate that in such a case the Hamiltonian system is
degenerate, $\det ||\partial H(J)/\partial J_{i}\partial J_{j}||=0$, and,
therefore, the Hamiltonian does not depend on some combinations of the
action variables. For example, suppose the integral $Y$ does not commute
with the integral $J_{N}$ only. Then the Hamiltonian can depend on the
variables $J_{1},\dots ,J_{N-1}$ only, otherwise $H\;$cannot commute with $Y$
. Thus, we see that the noncommutative algebra of integrals of motion allows
one to find canonical variables such that part of the corresponding action
variables disappears from the Hamiltonian. This phenomenon is closely
related to the topological properties of orbits for the Hamiltonian system.
Namely, trajectories of the integrable Hamiltonian system with $N$%
-dimensional commutative set of integrals of motion form (in the compact
case) a winding of $N$-torus in $2N$-dimensional phase space. If the set of
the integrals is noncommutative then the dimension of the corresponding
torus is $r<N$ (see \cite{Fomen83}).

The phenomenon of variable ''reduction'' takes place in the quantum
integrable Hamiltonian system as well. As will be demonstrated below, by
constructing a special isomorphism of linear functional spaces, we can
transform the initial differential operator of an equation into another one
with a reduced number of variables. The method which we are going to use for
such a demonstration, is, in fact, the harmonic analysis on the
noncommutative functional algebras.

Suppose a differential equation 
\begin{equation}
H(x,\partial _{x})\psi (x)=0,\quad \psi \in {\cal L}\subset C^{\infty
}(R^{N})  \label{e:1}
\end{equation}
with $N$ independent variables $x$ $\in R^{N}$ admits a noncommutative
algebra of functionally independent symmetry operators ${\cal F}%
=\{X_{a}(x,\partial _{x})\}$. The corresponding commutation relations are in
the general case nonlinear, 
\begin{equation}
\frac{i}{\hbar }[X_{a},X_{b}]=\Omega _{ab}(X),\quad a,b=1,\dots ,n\equiv
\dim {\cal F}..  \label{e:2}
\end{equation}
Here $\Omega _{ab}(X)$ are symmetric operator functions. The linear case,
when $\Omega _{ab}(X)=C_{ab}^{c}X_{c},$ corresponds to a Lie algebra, the
quadratic symmetric functions $\Omega _{ab}(X)$ \ correspond to a quadratic
algebra, and so on. The algebra ${\cal F}$\ $\ $corresponds to the algebra $%
{\cal F}^{\prime }=\{Y_{\alpha }(x,\partial _{x})\}$ of the invariant
operators on ${\cal L}$: 
\begin{equation}
\lbrack X_{a},Y_{\alpha }]=0,\quad \frac{i}{\hbar }[Y_{\alpha },Y_{\beta
}]=\omega _{\alpha \beta }(Y),\quad \alpha ,\beta =1,\dots ,n^{\prime
}\equiv \dim {\cal F}^{\prime }.  \label{e:3}
\end{equation}
We denote via $E({\cal F})$ and $E({\cal F}^{\prime })$ enveloping fields
for the algebras ${\cal F}$ and ${\cal F}^{\prime }$ respectively. Elements
of $E({\cal F})$ and $E({\cal F}^{\prime })$ are symmetrized operator
functions of the generating operators $X_{a}$ $Y_{\alpha }$. It is clear
that the centers of the enveloping fields coincide, i.e. $Z(E({\cal F}))=Z(E(%
{\cal F}^{\prime }))$. \ The elements of the center $Z=Z(E({\cal F}))$ are
called Casimir operators. The number of the independent Casimir operators,
which generate the center $Z$ is called the index of the algebra ${\cal F}$: 
$r\equiv {\rm ind}\,{\cal F}={\rm ind}\,{\cal F}^{\prime }$. If we replace
the operators $X$ and $Y$ \ in the operator functions $\omega _{\alpha \beta
}(Y)$ and $\Omega _{ab}(X)$ $\ $by arbitrary complex numbers $\xi $ $\ $and $%
f$, \ then the index of the algebras ${\cal F}$ and ${\cal F}^{\prime }$ can
be calculated according to the formula 
\begin{equation}
r=\sup_{\xi \in C}{\rm corank}\,\Omega _{ab}(\xi )=\sup_{f\in C}{\rm corank}%
\,\omega _{\alpha \beta }(f)\,.  \label{e:4}
\end{equation}
One can show that the following relation takes place 
\begin{equation}
n+n^{\prime }=2N.  \label{e:5}
\end{equation}

Let us introduce the notion of the\ $\lambda $-representation of the algebra 
${\cal F}$ \cite{Shiro00}. In fact, the\ $\lambda $-representation is the
result of the quantization of the classical Poisson bracket and can be
understood as a realization of the algebra ${\cal F}$ \ by an irreducible
set of differential operators $\tilde{X}=\tilde{X}(q,\partial _{q},j),$
dependent on $r$ parameters $j=(j_{1},\dots ,j_{r}),$ and acting in a space
of functions of$\ [q]=(n-r)/2$ independent variables\footnote{%
Via $\left[ q\right] $ we denote the number of the variables $q.$ Similar
notations are used below.} $q\in Q$ , 
\begin{equation}
\frac{i}{\hbar }[\tilde{X}_{a},\tilde{X}_{b}]=-\Omega _{ab}(\tilde{X}),\quad
K_{\mu }(\tilde{X}(q,\partial _{q},j))=\kappa _{\mu }(j),\quad \det \left|
\left| \frac{\partial \kappa _{\mu }(j)}{\partial j_{\nu }}\right| \right|
\neq 0\,.  \label{e:6}
\end{equation}
Here $K_{\mu }$ are all the independent Casimir operators of ${\cal F}$. In
a similar manner, we construct the $\lambda $-representation $\{\tilde{Y}\}$
of ${\cal F}^{\prime }\,$in a space of functions of$\ [q^{\prime
}]=(n^{\prime }-r)/2$ independent variables $q^{\prime }\in Q^{\prime }$ , 
\begin{equation}
\frac{i}{\hbar }[\tilde{Y}_{\alpha },\tilde{Y}_{\beta }]=\omega _{\alpha
\beta }(\tilde{Y}),\quad K_{\mu }^{\prime }(\tilde{Y}(q^{\prime },\partial
_{q^{\prime }},j))=K_{\mu }(\tilde{X}(q,\partial _{q},j))=\kappa _{\mu
}(j)\,.  \label{e:7}
\end{equation}
Suppose that in the spaces of functions of $x,$ of $q,$ and of $q^{\prime }$
are defined scalar products 
\begin{equation}
(\varphi ,\psi )=\int\limits_{R^{N}}\overline{\varphi (x)}\psi (x)d\mu
(x),\;(\tilde{\varphi},\tilde{\psi})=\int\limits_{Q}\overline{\tilde{\varphi}%
(q)}\tilde{\psi}(q)d\mu (q),\;(\tilde{\varphi},\tilde{\psi})^{\prime
}=\int\limits_{Q^{\prime }}\overline{\tilde{\varphi}(q^{\prime })}\tilde{\psi%
}(q^{\prime })d\mu (q^{\prime }),  \label{e:8}
\end{equation}
where $d\mu (x),$ $d\mu (q),\;$and $d\mu (q^{\prime })$ are some measures on 
$R^{N},$ $Q,\;$and $Q^{\prime }$ respectively. And suppose that the
operators $X_{a}(x,\partial _{x})$, $Y_{\alpha }(x,\partial _{x})$ and the
operators $\tilde{X}_{a}(q,\partial _{q},j),\tilde{Y}(q^{\prime },\partial
_{q^{\prime }},j)$ are self-conjugate with respect to the corresponding
scalar products\footnote{%
This supposition is not necessary and is introduced to simplify the
consideration.}. Now we define the set of distributions $D_{qq^{\prime
}}^{j}(x)$ as a solution of the overdetermined system of the equations: 
\begin{equation}
\left[ X_{a}(x,\partial _{x})-\tilde{X}_{a}(q,\partial _{q},j)\right]
D_{qq^{\prime }}^{j}(x)=0;\;\left[ Y_{\alpha }(x,\partial _{x})-\tilde{Y}%
_{a}(q^{\prime },\partial _{q^{\prime }},j)\right] D_{qq^{\prime
}}^{j}(x)=0\,.  \label{e:9}
\end{equation}
The distributions $D_{qq^{\prime }}^{j}(x)$ obey the completeness and
orthogonality relations: 
\begin{equation}
\int \overline{D_{qq^{\prime }}^{j}(x)}D_{\tilde{q}\tilde{q}^{\prime }}^{%
\tilde{j}}(x)\,d\mu (x)=\delta (j,\tilde{j})\delta (q,\tilde{q})\delta
(q^{\prime },\tilde{q}^{\prime });  \label{e:11}
\end{equation}
\begin{equation}
\int \overline{D_{qq^{\prime }}^{j}(x)}D_{qq^{\prime }}^{j}(\tilde{x})\,d\mu
(j)d\mu (q)d\mu (q^{\prime })=\delta (x,\tilde{x}).  \label{e:12}
\end{equation}
Here $d\mu (j)$ is the spectral measure of the Casimir operators $K(X)\
(=K^{\prime }(Y))$. Due to Eqs.\ (\ref{e:6}), (\ref{e:7}) (\ref{e:9}) the
distributions $D_{qq^{\prime }}^{j}(x)$ are eigenfunctions of all the
Casimir operators,

\begin{equation}
K_{\mu }(X(x,\partial _{x}))D_{qq^{\prime }}^{j}(x)=\kappa _{\mu
}(j)D_{qq^{\prime }}^{j}(x),\quad \mu =1,\dots ,r\,.  \label{e:13}
\end{equation}
Usually one can find $D_{qq^{\prime }}^{j}(x)$ by an integration, at least
in the case when ${\cal F}$ is a Lie algebra. As a consequence of (\ref{e:12}%
) and (\ref{e:11}) we can define the direct and the inverse Fourier
transforms: 
\begin{eqnarray}
&&\tilde{\psi}(q,q^{\prime },j)=\int D_{qq^{\prime }}^{j}(x)\overline{\psi
(x)}\,d\mu (x)\,,  \label{e:14} \\
&&\psi (x)=\int D_{qq^{\prime }}^{j}(x)\overline{\tilde{\psi}(q,q^{\prime
},j)}d\mu (j)d\mu (q)d\mu (q^{\prime })\,.  \label{e:15}
\end{eqnarray}
The Eqs. (\ref{e:14}) and (\ref{e:15}) establish an isomorphism of the
spaces ${\cal L}$ and $\tilde{{\cal L}}=\{\tilde{\psi}\}.$ It is important
to stress that under such an isomorphism the operators $X$, $Y$ are
transformed into the differential operators $\tilde{X},\tilde{Y}$ that act
in the spaces of functions, which depend on a fewer number of variables, 
\begin{equation}
X(x,\partial _{x})\psi (x)\leftrightarrow \tilde{X}(q,\partial _{q},j)\tilde{%
\psi}(q,q^{\prime },j),\quad Y(x,\partial _{x})\psi (x)\leftrightarrow 
\tilde{Y}(q^{\prime },\partial _{q^{\prime }},j)\tilde{\psi}(q,q^{\prime
},j).  \label{e:16}
\end{equation}

Let us return to the equation (\ref{e:1}). Here we can conclude that $H\in E(%
{\cal F}^{\prime })$ since the operator $H$ commutes with all the operators
of the algebra ${\cal F}=\{X_{a}\}$. In turn, that means that there exists
an operator function $H(Y)$ such that $H(x,\partial _{x})=H(Y(x,\partial
_{x}))$. Let us look for solutions of the equation (\ref{e:1}) in the form (%
\ref{e:15}). Using the isomorphism (\ref{e:16}), we get: 
\begin{equation}
H(\tilde{Y}(q^{\prime },\partial _{q^{\prime }},j))\tilde{\psi}(q,q^{\prime
},j)=0.  \label{e:17}
\end{equation}
Thus, departing from the equation (\ref{e:1}) without loss of any
information, we have arrived to a differential equation with $\tilde{N}%
=[q^{\prime }]=(n^{\prime }-r)/2$ independent variables. Taking into account
(\ref{e:5}), one can obtain 
\begin{equation}
\tilde{N}=N-\frac{1}{2}(\dim {\cal F}+{\rm ind}\,{\cal F}).  \label{e:18}
\end{equation}

Thus, the existence of a noncommutative symmetry algebra results into the
phenomenon of variable reduction. Indeed, we started with $N=[q^{\prime
}]+[q]+r$ variables. $[q]$ variables have disappeared completely, and $r=[j]$
variables remain in the equations as some parameters. Respectively, the
solution of the equation (\ref{e:17}) contains as a factor an arbitrary
function of the variables $q,j$. The number of variables $q$ is equal to $%
[q]=(n-r)/2=(\dim {\cal F}-{\rm ind}\,{\cal F})/2$. In the commutative
case:\ ${\rm ind}\,{\cal F}=\dim {\cal F}$, \ and $[q]=0$. In the
noncommutative case: $\dim {\cal F}>{\rm ind}\,{\cal F}$, \ and $[q]>0$.
Thus, the reduction of variables always takes place when there exists a
noncommutative algebra of the integrals of motion.

Let us apply the above consideration to the Klein-Gordon equation (\ref{2.2}%
) in a uniform magnetic field. In this case we have four\ ($N=4$) variables
and five $(n=5)$ independent symmetry operators: ${\cal F}%
=\{P_{0},P_{3},a_{2},a_{2}^{+},L_{z}=\hbar L\},$ 
\[
P_{0}=i\hbar \partial _{0},\;P_{3}=i\hbar \partial _{3},\;L=u\partial _{u}-{%
\bar{u}}\partial _{{\bar{u}}}\,,\;a_{2}=\partial _{{\bar{u}}%
}+u/2,\;a_{2}^{+}=-\partial _{u}+{\bar{u}}/2\,,\;u\equiv x+iy\,. 
\]
The non-zero commutation relations are $[a_{2},a_{2}^{+}]=1,\quad \lbrack
L,a_{2}]=a_{2},\quad \lbrack L,a_{2}^{+}]=-a_{2}^{+}.$ It follows from (\ref
{e:4}) that $r={\rm ind}\,{\cal F}=3$. $\tilde{N}=[q^{\prime }]=0$ ${\cal K}%
\in Z$ according to (\ref{e:18}). Thus, the equations (\ref{e:17}) present
algebraic relations on the parameters $j$ (and on the parameters of the
equation itself ). Besides, $\dim {\cal F}^{\prime }=3=r$,\ due to (\ref{e:5}%
). Thus, the algebra of the invariant operators is placed completely in the
center. Or more simply, there are no operators $Y,\tilde{Y}$ \ and
variables\ $q^{\prime }$ in the case under consideration. The center $Z$ is
generated by three Casimir operators, those are $K_{1}=P_{0},\;K_{2}=P_{3},%
\;K_{3}={\cal N}=L+a_{2}^{+}a_{2}=L+\frac{1}{2}%
(a_{2}^{+}a_{2}+a_{2}a_{2}^{+}-1)$.

Let us construct the $\lambda $-representation of the algebra ${\cal F}$: 
\[
\tilde{P}_{0}=j_{1}=\hbar k_{0}\,,\;\tilde{P}_{3}=j_{2}=\hbar k_{3}\,,\;%
\tilde{a}_{2}=q\,,\;\tilde{a}_{2}^{+}=\partial _{q}\,,\;\tilde{L}=-q\partial
_{q}+n\,,\;n=j_{3}=0,1,\dots \;. 
\]
The operators $\tilde{a}_{2}$ and $\tilde{a}_{2}^{+}$ \ are mutually
conjugate with respect to the scalar product (\ref{e:8}) with the measure $%
d\mu (q)=\exp (-q\bar{q})d^{2}q/\pi ,\;\ (d^{2}q\equiv
dq_{1}dq_{2},\;q=q_{1}+iq_{2})$. The operator $\tilde{L}$ is self-conjugate,
and the space $\tilde{{\cal L}}$ is built up from analytic functions
dependent on the variable $q$ and on the parameters $j=(k_{0},k_{3},n)$.
Here the Casimir operator $\tilde{{\cal N}}$ \ has the following form \ $%
\tilde{{\cal N}}=\tilde{L}+\frac{1}{2}(\tilde{a}_{2}^{+}\tilde{a}_{2}+ 
\tilde{a}_{2}\tilde{a}_{2}^{+}-1)=n$.

From the equations (\ref{e:9}) we can find the set $D_{q}^{j}(x)$ that obeys
the completeness and orthogonality relations. Such a set has the form 
\begin{eqnarray}
&&D_{q}^{j}(x)=e^{i(k_{0}x^{0}+k_{3}x^{3})}e^{q{\bar{u}}-u{\bar{u}}%
/2}(u-q)^{n}/(2\pi \sqrt{\pi n!})\,,  \label{e:19} \\
&&\int \overline{D_{q}^{j}(x)}D_{\tilde{q}}^{\tilde{j}}(x)\,dx=\delta (k_{0}-%
\tilde{k}_{0})\delta (k_{3}-\tilde{k}_{3})\delta (\tilde{q},q)\delta _{n%
\tilde{n}}\,,  \label{e:20} \\
&&\sum\limits_{n=0}^{\infty }\int \overline{D_{q}^{j}(x)}D_{q}^{j}(\tilde{x}%
)\,dk_{0}dk_{3}d\mu (q)=\delta (x^{0}-\tilde{x}^{0})\delta (x^{3}-\tilde{x}%
^{3})\delta (x-\tilde{x})\delta (y-\tilde{y}).  \label{e:21}
\end{eqnarray}
In Eq.(\ref{e:20}) $\delta (\tilde{q},q)=\exp (\bar{q}{\tilde{q}})$ is a $%
\delta $-function with respect to the measure $d\mu (q)$. To justify the
validity of (\ref{e:20}) and (\ref{e:21}) one can use the following
relations \cite{Perel87} 
\[
\int \overline{v_{n}(q)}v_{m}(q)\,d\mu (q)=\delta
_{nm},\;\sum\limits_{n=0}^{\infty }\overline{v_{n}(q)}v_{n}(\tilde{q}%
)=\delta (\tilde{q},q)=\exp (\bar{q}{\tilde{q}})\,,\;v_{n}(q)\equiv q^{n}/%
\sqrt{n!}\,. 
\]

As was already mentioned above, the Klein-Gordon operator ${\cal K}$ belongs
to the center $Z$ , and , therefore, can be presented in a polynomial form
of the Casimir operators $P_{0},P_{3},{\cal N}$, which generate this center.
Such a representation is given by Eq. (\ref{2.12}). The Klein-Gordon
equation in the space $\tilde{{\cal L}}$, i.e. the equation (\ref{e:17}) is,
in fact, the relation (\ref{3.3}) for the parameters $j=(k_{0},k_{3},n)$.
Thus, the functions (\ref{e:19}) form a basis of the Klein-Gordon equation\ (%
\ref{2.2}) (with allowance made for (\ref{3.3}).

\begin{acknowledgement}
V.G.B thanks FAPESP for support and Nuclear Physics Department of S\~{a}o
Paulo University for hospitality, as well as he thanks Russian Science
Ministry Foundation and RFFI for partial support; M.C.B. thanks FAPESP and
D.M.G thanks both FAPESP and CNPq for permanent support.
\end{acknowledgement}

\end{document}